\documentclass[useAMS,usenatbib]{mn2e}

\usepackage{epsfig}
\usepackage{longtable}
\usepackage{times} 
\newcommand{\sw}{$Swift$}

\def \sw {{\em Swift}}

\def \hcm {\hbox {\ifmmode $ atom cm$^{-2}\else atom cm$^{-2}$\fi}}

\def \ATel {Astron.\ Tel.}
\def \apj {ApJ}
\def \apjl {ApJL}
\def \apjs {ApJS}\def \aap {A\&A}

\def \pasj {PASJ}

\def \mnras {MNRAS}

\title[SFXTs with {\it Swift}: the first year]{Monitoring Supergiant Fast X--ray Transients with \emph {Swift}. \\
 Results from the first year }

\author[P.\ Romano et al.]{P.\ Romano$^{1}$, L.\ Sidoli$^{2}$, G.\ Cusumano$^{1}$, V.\ La Parola$^{1}$, S.\ Vercellone$^{1}$, C.\ Pagani$^{3}$, 
\newauthor L.\ Ducci$^{4,2}$, V.\ Mangano$^{1}$, J.\ Cummings$^{5}$, H.A.\ Krimm$^{5,6}$, C.\ Guidorzi$^{7}$, J.A.~Kennea$^{3}$, 
 \newauthor  E.A.~Hoversten$^{3}$, D.N.~Burrows$^{3}$, N.~Gehrels$^{5}$ \\
$^{1}$INAF, Istituto di Astrofisica Spaziale e Fisica Cosmica,
        Via U.\ La Malfa 153, I-90146 Palermo, Italy\\
$^{2}$INAF, Istituto di Astrofisica Spaziale e Fisica Cosmica,
	Via E.\ Bassini 15,   I-20133 Milano,  Italy\\
$^{3}$Department of Astronomy and Astrophysics, Pennsylvania State 
             University, University Park, PA 16802, USA\\
$^{4}$Dipartimento di Fisica e Matematica, Universit\`a dell'Insubria, Via
Valleggio 11, I-22100 Como, Italy \\
$^{5}$NASA/Goddard Space Flight Center, Greenbelt, MD 20771, USA\\
$^{6}$Universities Space Research Association, Columbia, MD, USA \\
$^{7}$Dipartimento di Fisica, Universit\`a{} di Ferrara, Via Saragat 1, I-44100 Ferrara, Italy \\
}

\begin{document}

\vspace{-2truecm}

\date{Accepted 2009 July 07. Received 2009 July 07; in original form 2009 May 24}

\pagerange{\pageref{firstpage}--\pageref{lastpage}} \pubyear{2009}

\maketitle

\label{firstpage}

\begin{abstract}
The advent of {\it Swift} has allowed, for the first time, the possibility to
give Supergiant Fast X--ray Transients (SFXTs), the new class of High Mass
X--ray Binaries discovered by INTEGRAL, non serendipitous attention 
throughout most phases of their life. 
In this paper we present our results based on the first year of intense \sw\ 
monitoring of four SFXTs, IGR~J16479$-$4514,  XTE~J1739$-$302, IGR~J17544$-$2619,
and  AX~J1841.0$-$0536. 

We obtain the first assessment of how long each source 
spends in each state using a systematic monitoring with a sensitive instrument.
The duty-cycle of inactivity is  $\sim 17, 28, 39, 55$\,\%  ($\sim $5\,\% uncertainty),  
for IGR~J16479$-$4514, AX~J1841.0$-$0536, XTE~J1739--302, and IGR~J17544$-$2619,  
respectively, so that true quiescence, 
which is below our detection ability even with the 
exposures we collected in one year, is a rare state, 
when compared  with estimates from less sensitive instruments. 
This demonstrates that these transients accrete matter throughout their lifetime at 
different rates.  

AX~J1841.0$-$0536 is the only source which has not undergone a bright outburst
during our monitoring campaign. 
Although individual sources behave somewhat differently, 
common X--ray characteristics of this class are 
emerging such as outburst lengths well in excess of hours, with a
multiple peaked structure. A high dynamic range (including bright outbursts) 
of $\sim4$ orders of magnitude have been observed
in IGR~J17544$-$2619 and  XTE~J1739$-$302, of $\sim$3 in  IGR~J16479$-$4514, and of about 2 in 
AX~J1841.0$-$0536 (this lowest range is due to the lack of bright flares). 
We also present a complete list of BAT on-board detections, which complements 
our previous work, and further confirms the continuous activity of these sources. 

We performed out-of-outburst intensity-based spectroscopy. 
In particular, spectral fits with an absorbed blackbody always result in blackbody 
radii of a few hundred meters, consistent with being emitted from a small portion 
of the neutron star surface, very likely the neutron star polar caps. 

We used the whole BAT dataset, since the beginning of the mission, 
to search for periodicities due to orbital motion and found 
$P_{\rm orb}=3.32$\,d for IGR~J16479$-$4514, confirming previous findings. 
We also present the UVOT data of these sources; we show the UVOT light curves 
of AX~J1841.0$-$0536 and the ones of XTE~J1739$-$302 before, during, and after the outbursts.  

\end{abstract}

\begin{keywords}
X-rays: binaries -- X-rays: individual: IGR~J16479$-$4514, XTE~J1739--302, 
IGR~J17544$-$2619,  AX~J1841.0$-$0536. 

\noindent
Facility: {\it Swift}

\end{keywords}


	\section{Introduction\label{sfxt5:intro}}

Supergiant Fast X--ray Transients (SFXTs) are a sub-class of High Mass
X--ray Binaries (HMXBs)
recently discovered by INTEGRAL during the Galactic Plane monitoring \citep{Sguera2005}. 
They are firmly associated (via optical spectroscopy) with an O or B supergiant and 
display outbursts which are significantly shorter than typical
Be/X-ray binaries, characterized by bright flares with a duration
of a few hours and peak luminosities of 10$^{36}$--10$^{37}$~erg~s$^{-1}$. 
The quiescence, characterized by a soft spectrum (likely thermal) and a 
low luminosity at $\sim 10^{32}$~erg~s$^{-1}$ is a rarely-observed state
\citep[e.g.][]{zand2005}. 
As their spectral properties resemble those of accreting
pulsars, it is generally assumed that all members of the new class are 
HMXBs hosting a neutron star, although the only 
three SFXTs with a measured pulse period are  
AX~J1841.0$-$0536 ($P_{\rm spin}\sim4.7$\,s, \citealt{Bamba2001}), IGR~J11215--5952
($P_{\rm spin}\sim187$\,s, \citealt{Swank2007}), and 
IGR~J18483--0311 ($P_{\rm spin}\sim21$\,s, \citealt{Sguera2007}).
The mechanisms responsible for the observed short outbursts
are still being debated. The proposed explanations 
(see \citealt{Sidoli2009:cospar}, for a recent review) 
mainly involve the structure of the wind from the supergiant companion 
\citep{zand2005,Walter2007,Negueruela2008,Sidoli2007}, or the possible
presence of gated mechanisms \citep[see,][]{Bozzo2008}. The latter are 
due to the properties of the
accreting neutron star (magnetar-like magnetic fields and slow pulse periods) 
which can halt the accretion for most of the time.

During February 2007, we monitored the outburst of the periodic SFXT IGR~J11215$-$5952 
\citep{Romano2007,Sidoli2007} with {\it Swift} \citep{Gehrels2004mn}, in what became 
the most complete and deep set of X--ray observations of an outburst of a SFXT. 
Thanks to these observations, we discovered that the accretion phase during the 
bright outburst lasts much longer than 
a few hours. The orbital dependence of the accretion X--ray luminosity during the outburst
led us to propose an alternative explanation 
for the outburst mechanism in IGR~J11215$-$5952 \citep{Sidoli2007}, 
linked to the possible presence of a second wind component, 
in the form of a preferred plane for the outflowing wind from the supergiant donor.
X--ray outbursts  should be produced 
when the neutron star crosses this density enhanced wind component.
%


 \begin{table*}
 \begin{center}
 \caption{Summary of the {\it Swift}/XRT monitoring campaign of the four SFXTs during the first year.\label{sfxt5:tab:campaign} }
 \begin{tabular}{lrrrrlll}
 \hline
 \noalign{\smallskip}
Name &Campaign &Campaign &N$^{\mathrm{a}}$ &Exposure$^{\mathrm{b}}$ &Outburst$^{\mathrm{c}}$ &BAT &References \\
     &       Start         &End             & &       & Dates   & Trigger   & \\
     &       (yyyy-mm-dd)  &(yyyy-mm-dd)    & & (ks)  & (yyyy-mm-dd) &   & \\
  \noalign{\smallskip}
 \hline
 \noalign{\smallskip}
IGR~J16479$-$4514 &  2007-10-26 & 2008-10-25& 70&  75.2 & 2008-03-19&       306829       &\citet{Romano2008:sfxts_paperII}\\
                  &             &           &   &       & 2008-05-21&       312068       & \\ 
                  &   	        &           &   &       & {\it 2009-01-29}& {\it 341452} & {\it \citet{Romano2009:atel1920,LaParola2009:atel1929} } \\ 
XTE~J1739$-$302   &  2007-10-27 & 2008-10-31& 95&  116.1& 2008-04-08&       308797       & \citet{Sidoli2009:sfxts_paperIII}\\
                  &   	        &           &   &       & 2008-08-13&       319963       & \citet{Romano2008:atel1659}, \citet{Sidoli2009:sfxts_paperIV} \\
                  &   	        &           &   &       & {\it 2009-03-10}& {\it 346069} & {\it \citet{Romano2009:atel1961} }\\ 
IGR~J17544$-$2619 &  2007-10-28 & 2008-10-31& 77&  74.8 & 2007-11-08&                    & \citet{Krimm2007:ATel1265}\\
  		  &    	&           &   &       & 2008-03-31&       308224       & \citet{Sidoli2009:sfxts_paperIII}\\
  		  &   	        &           &   &       & 2008-09-04&                    & \citet{Romano2008:atel1697,Sidoli2009:sfxts_paperIV}   \\
                  &   	        &           &   &       & {\it 2009-03-15}&              &{\it \citet{Krimm2009:atel1971} }\\ 
AX~J1841.0$-$0536 &  2007-10-26 & 2008-11-15& 88&  96.5 & none      &                    &  \\  
  \noalign{\smallskip}
 \hline
 \noalign{\smallskip}
Total             &             &           &330&  362.6 &           &              & \\  
  \noalign{\smallskip}
  \hline
  \end{tabular}
  \end{center}
  \begin{list}{}{}
  \item[$^{\mathrm{a}}$]{Number of observations obtained during the monitoring campaign.}
  \item[$^{\mathrm{b}}$]{\sw/XRT net exposure.}
  \item[$^{\mathrm{c}}$]{BAT trigger dates. We report the outburst that occurred in 2009 in italics, for the sake of completeness. }
  \end{list}
  \end{table*}


Following the success of the {\it Swift} observations on IGR~J11215--5952, 
we extended the investigation to a small although well-defined 
sample of SFXTs.
{\it Swift} was the most logical choice to monitor the light curves of our sample,
because of its unique fast-slewing and flexible observing scheduling,
which makes a monitoring effort cost-effective, 
its broad-band energy coverage that would allow us to model the 
observed spectra simultaneously in the 0.3--150\,keV energy range,
thus testing the prevailing models for accreting neutron stars,
and the high sensitivity in the soft X-ray regime, where some of the
SFXTs had never been observed.

In \citet[][Paper I]{Sidoli2008:sfxts_paperI}, we described the 
long-term X--ray emission outside the bright outbursts based on the first 4 months of data; 
in \citet[][Paper II]{Romano2008:sfxts_paperII} and 
\citet[][Paper III]{Sidoli2009:sfxts_paperIII}, 
we reported on the outbursts of IGR~J16479$-$4514,
and the prototypical IGR~J17544$-$2619 and XTE~J17391$-$302, respectively, 
while in \citet[][Paper IV]{Sidoli2009:sfxts_paperIV} we report the 
results of more outbursts of XTE~J1739$-$302 and IGR~J17544$-$2619. 
In this paper we draw a general picture of our knowledge on SFXT, by summarizing 
the results on the outbursts caught by \sw\ during the first year of our 
ongoing campaign, and report on the XRT \citep{Burrows2005:XRTmn} 
and UVOT \citep{Roming2005:UVOTmn}   
data collected from 2007 October 26 to  2008 November 15, 
as well as the BAT \citep{Barthelmy2005:BAT} data 
collected since the start of the mission. 
We also include data from the 2009 January 29 outburst of IGR~J16479$-$4514
\citep{Romano2009:atel1920}.

 \begin{table*}
 \begin{center}
 \caption{Duty cycle of inactivity of the four SFXTs.}
 \label{sfxt5:tab:dutycycle}
 \begin{tabular}{lcccccccc}
 \hline
 \noalign{\smallskip}
Name &  Limiting Rate$^{\mathrm{a}}$ & Limiting $F$$^{\mathrm{a,b}}$  & Limiting $L$$^{\mathrm{a,b}}$ &$\Delta T_{\Sigma}$  & $P_{\rm short}$ &  IDC 
                  & Rate$_{\Delta T_{\Sigma}}$   \\     
               & (0.2--10\,keV)   &  (2--10\,keV) &  (2--10\,keV)  &  &  &  & (0.2--10\,keV)       \\                  
                & ($10^{-3}$ counts s$^{-1}$) &  ($10^{-12}$ erg cm$^{-2}$ s$^{-1}$)  &($10^{35}$ erg s$^{-1}$)
                &(ks) & (\%) &  (\%) &   ($10^{-3}$counts s$^{-1}$)
                   \\  
  \noalign{\smallskip}
 \hline
 \noalign{\smallskip}
IGR~J16479$-$4514   &16  &2.4  &0.62& 12.2 &2  & 17  & $2.9\pm0.7$        \\
XTE~J1739$-$302     &13  &1.6  &0.13& 40.3 &9  & 39  & $3.9\pm0.4$       \\
IGR~J17544$-$2619   &13  &1.2  &0.17& 37.0 &10 & 55  & $1.9\pm0.3$       \\
AX~J1841.0$-$0536   &13  &1.7  &0.45& 26.6 &3  & 28  & $2.4\pm0.4$    \\     
  \noalign{\smallskip}
  \hline
  \end{tabular}
  \end{center}
  \begin{list}{}{}
\item {Count rates are in units of $10^{-3}$ counts s$^{-1}$ in the 0.2--10\,keV energy band.  
         Observed fluxes and luminosities are in units of $10^{-12}$ erg cm$^{-2}$ s$^{-1}$ 
         and $10^{35}$ erg s$^{-1}$ in the 2--10\,keV energy band,
         respectively. $\Delta T_{\Sigma}$ is sum of the exposures accumulated in all observations, 
   each in excess of 900\,s, where only a 3-$\sigma$ upper limit was achieved;  
   $P_{\rm short}$ is the percentage of time lost to short observations; 
   IDC is the  {\it duty cycle of inactivity}, 
   the time each source spends  undetected down to a flux limit of 1--3$\times10^{-12}$ erg cm$^{-2}$ s$^{-1}$;
   Rate$_{\Delta T_{\Sigma}}$ is detailed in the text (Sect.~\ref{sfxt5:idc}).
   }
  \item[$^{\mathrm{a}}$]{Based on a single 900\,s exposure.}
  \item[$^{\mathrm{b}}$]{Based on the best fit model for the `low' (or `medium' if `low' unavailable)
      absorbed power-law model in Table~\ref{sfxt5:tab:specfits}.} 
  \end{list}
  \end{table*}


	\section{Our sample and Observations\label{sfxt5:sample}}

The four targets, IGR~J16479$-$4514, XTE~J1739--302, 
IGR~J17544$-$2619, and AX~J1841.0$-$0536 
were selected by considering sources which, among several SFXT candidates, are
confirmed SFXTs, i.e.\ they display both a `short' transient (and recurrent) 
X--ray activity and they have been optically identified with
supergiant companions [see \cite{Walter2007} and references therein].  
XTE~J1739--302 and IGR~J17544$-$2619, in particular, are generally 
considered prototypical SFXTs: XTE~J1739--302 was the first transient which showed
an unusual X--ray behavior \citep{Smith1998:17391-3021}, 
only recently optically associated with a blue supergiant \citep{Negueruela2006}.
AX~J1841.0$-$0536/IGR~J18410$-$0535, was chosen because at the time it was the only
SFXT, together with IGR~J11215--5952, where a pulsar had been detected \citep{Bamba2001}.
Finally, IGR~J16479$-$4514 had displayed in the past a more frequent X--ray 
outburst occurrence than other SFXTs \citep[see, e.g.][]{Walter2007}, 
and offered an {\it a priori} better chance to be caught during an outburst.   

For these sources we obtained 2--3 observations week$^{-1}$ object$^{-1}$, 
each 1\,ks long with XRT  in AUTO mode, 
to best exploit XRT automatic mode switching \citep{Hill04:xrtmodes_mn} 
in response to changes in the observed fluxes. 
This observing pace would naturally fit in the regular observation scheduling 
of $\gamma$-ray bursts (GRBs), which are the main observing targets for 
{\it Swift}. 
We also proposed for further target of opportunity (ToO)
observations whenever one of the sources showed interesting activity,
(such as indications of an imminent outburst)  or
underwent an outburst, thus obtaining a finer sampling of the light
curves and allowing us to study all phases of the evolution of an
outburst. 

During the first year, we collected a total of 330 \sw\ observations as part of our program, for a 
total net XRT exposure of $\sim 363$\,ks accumulated on all sources and 
distributed as shown in Table~\ref{sfxt5:tab:campaign}. 

In this paper we also include the data on the 20\,d campaign 
(for a total on-source time of $\sim 34$\,ks) on 
IGR~J16479$-$4514, which triggered the BAT on 2009 January 29 at 06:33:07 UT
\citep[image trigger=341452,][]{Romano2009:atel1920}. 
\sw\ slewed to the target so that the XRT started observing the
field at 06:46:46.9 UT, 819.3\,s after the BAT trigger. 
The BAT transient monitor showed enhanced emission (in excess of 20 mCrab) 
from 01:38:56 to 07:02:08 UT. 
During the image trigger interval (the 640 seconds starting at 2009-01-29 06:27:5) 
the rate was $0.022\pm0.003$ counts s$^{-1}$ (97 mCrab). 
IGR~J16479$-$4514 showed renewed activity on 2009 February 8, starting from about 20:30 UT
\citep{LaParola2009:atel1929}. 
For the 504\,s pointing starting at 2009-02-08 20:30 UT 
the BAT transient monitor rate was $0.019\pm0.003$ counts s$^{-1}$ (85 mCrab).

 	 \section{ Data Reduction\label{sfxt5:dataredu}}

The XRT data were uniformly processed with standard procedures 
({\sc xrtpipeline} v0.11.6), filtering and screening criteria by using 
{\sc FTOOLS} in the {\sc Heasoft} package (v.6.4).  
We considered both WT and PC data, 
and selected event grades 0--2 and 0--12, respectively 
(\citealt{Burrows2005:XRTmn}).
When appropriate, we corrected for pile-up 
by determining the size of the point spread function (PSF) core affected 
by comparing the observed and nominal PSF \citep{vaughan2006:050315mn},
and excluding from the analysis all the events that fell within that
region. 
We used the spectral redistribution matrices v010 in CALDB.

We retrieved the BAT orbit-by-orbit light curves (15--50\,keV) 
covering the data range from February 12, 2005 
to December 31, 2008 (MJD range 53413--54831) from the 
BAT Transient Monitor \citep[][]{Krimm2006_atel_BTM,Krimm2008_HEAD_BTM} 
page\footnote{http://swift.gsfc.nasa.gov/docs/swift/results/transients/ }.

The UVOT observed the 4 targets simultaneously with the XRT 
with the `Filter of the Day', i.e.\ the filter chosen for all observations 
to be carried out during a specific day in order to minimize the filter 
wheel usage. The only exception are the observations during outbursts, 
when all filters were used in the typical GRB sequence \citep{Roming2005:UVOTmn}. 
The data analysis was performed using the {\sc uvotimsum} and 
{\sc uvotsource} tasks included in the {\sc FTOOLS} software. The latter 
task calculates the magnitude through aperture photometry within
a circular region and applies specific corrections due to the detector
characteristics. The reported magnitudes are on the UVOT photometric 
system described in \citet{Poole2008:UVOTmn}, and 
are not corrected for Galactic extinction. 
At the position of IGR~J16479$-$4514, no detection
was achieved down to a limit of $u=21.07$ mag.
For IGR~J17544$-$2619 only engineering data were collected, 
as is generally the case for a field which contains a source too bright to be observed;
the only exceptions were the outburst segments 00308224000, and the two following it,
00035056021 and 00035056023 [see Table~\ref{sfxts:tab:alldata17544}, and \citet{Sidoli2009:sfxts_paperIII}],  
where we observe $v=12.8$ mag and $uvw2=18.13\pm0.05$ and $uvw2=18.00\pm0.06$ mag, 
respectively.

All quoted uncertainties are given at 90\,\% confidence level for 
one interesting parameter unless otherwise stated. 
The spectral indices are parameterized as  
$F_{\nu} \propto \nu^{-\alpha}$, 
where $F_{\nu}$ (erg cm$^{-2}$ s$^{-1}$ Hz$^{-1}$) is the 
flux density as a function of frequency $\nu$; 
we adopt $\Gamma = \alpha +1$ as the photon index, 
$N(E) \propto E^{-\Gamma}$ (ph cm$^{-2}$ s$^{-1}$ keV$^{-1}$).

\section{Timing}

\begin{figure}
\begin{center}
\centerline{\includegraphics[width=10cm,angle=0]{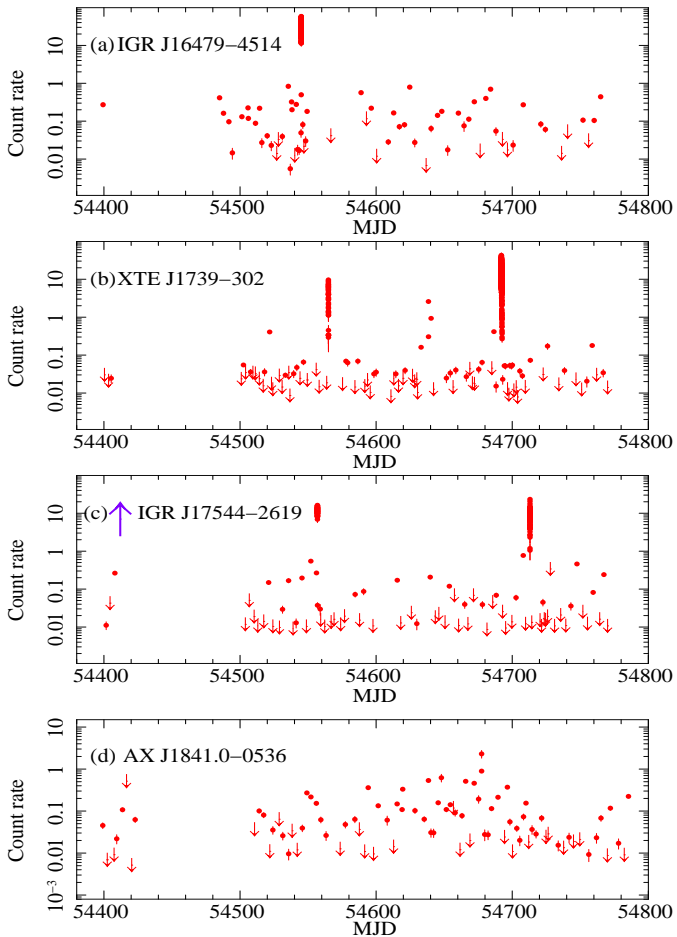}}
\vspace{-1.0truecm}
\caption[XRT light curves]{\sw/XRT (0.2--10\,keV) light curves, corrected for pile-up, 
                PSF losses, vignetting and background-subtracted. The data were 
                collected from 2007 October 26 to  2008 November 15. 
		The downward-pointing arrows are 3-$\sigma$ upper limits. The upward pointing arrow 
                marks an outburst that triggered the BAT on MJD 54,414, but which could not be followed by XRT
                because the source was Sun-constrained for the XRT. 
                }
		\label{sfxt5:fig:xrtlcvs} 
        \end{center}
        \end{figure}

        \subsection{XRT Inactivity Duty Cycle\label{sfxt5:idc}}

Fig.~\ref{sfxt5:fig:xrtlcvs} shows the XRT light curves 
collected from 2007 October 26 to  2008 November 15,
in the 0.2--10\,keV band, which were corrected for pile-up, PSF losses, and
vignetting, and background-subtracted. Each point in the
light curves refers to the average flux observed
during each observation performed with XRT; the exception are the outbursts 
(listed in Table~\ref{sfxt5:tab:campaign}) where the data were binned to 
include at least 20 source counts per time bin to best represent the 
count rate dynamical range. 
Due to the sources being Sun-constrained
between roughly 2007 December and 2008 January, depending on the target 
coordinates, no data were collected during those months. 

Given the structure of the observing plan we can realistically consider our monitoring as a 
casual sampling of the light curve at a resolution of about $\sim 4$\,d. 
Therefore, we can calculate the percentage of time each source spent in each relative 
flux state. 
In order to do this, we divided the observations into three states, namely 
{\it i)} BAT-detected outburst, 
{\it ii)} intermediate state (all observations yielding a firm detection excluding outburst ones),  
{\it iii)} `non detections' (detections with a significance below 3$\sigma$). 
Since a few observations were interrupted by GRB events, the consequent non detection may be 
due to the short exposure, not exclusively to the source being faint. 
Therefore, to create a uniform subsample 
for the latter state, we excluded all observations that had a net exposure below 900\,s.  
An exposure of 900\,s corresponds to 2--10\,keV flux limits that vary between 1 and 
3$\times 10^{-12}$ erg cm$^{-2}$ s$^{-1}$ (3$\sigma$), depending on the source. 
These values were derived from a measurement of the local background
and a count rate to flux conversion calculated by using the best fit absorbed power-law models of 
the `low' (or `medium' if `low' was not available) state in Table~\ref{sfxt5:tab:specfits}. 

We define as {\it duty cycle of inactivity}, 
the time each source spends {\it undetected} down to a flux limit of 
1--3$\times10^{-12}$ erg cm$^{-2}$ s$^{-1}$,  
$${\rm IDC}= \Delta T_{\Sigma} / [\Delta T_{\rm tot} \, (1-P_{\rm short}) ] \, , $$  where  
$\Delta T_{\Sigma}$ is sum of the exposures accumulated in all observations, 
   each in excess of 900\,s, where only a 3-$\sigma$ upper limit was achieved (Table~\ref{sfxt5:tab:dutycycle}, column 5), 
$\Delta T_{\rm tot}$ is the total exposure accumulated (Table~\ref{sfxt5:tab:campaign}, column 5), and 
$P_{\rm short}$ is the percentage of time lost to short observations 
   (exposure $<900$\,s, Table~\ref{sfxt5:tab:dutycycle}, column 6). 
We obtain that ${\rm IDC} \sim 17, 28, 39, 55$\,\%,  
for IGR~J16479$-$4514, AX~J1841.0$-$0536, XTE~J1739--302, and IGR~J17544$-$2619, respectively
(Table~\ref{sfxt5:tab:dutycycle}, column 7). 
We estimate an error of $\sim 5\,\%$ on these values.

We accumulated all data for which no detections were obtained as single exposures 
(whose total exposure is $\Delta T_{\Sigma}$), and performed a detection.  
The resulting cumulative mean count rate for each object is reported in 
Table~\ref{sfxt5:tab:dutycycle} (column 8).

       \subsection{BAT on-board detections}

In Table~\ref{sfxt5:tab:bat_triggers} we list the BAT on-board detections in the 15--50\,keV band.
If an alert was generated, a BAT trigger was assigned (cols. 4 and 8) and notices were 
disseminated\footnote{http://gcn.gsfc.nasa.gov/gcn/swift\_grbs.html}.  
For some of these triggers a burst response (slew and repointing of the NFI) was also initiated,
depending on GRB observing load, observing constraints, and interest in the sources.
More details on several of these triggers can be found in the papers of our series
(see Table~\ref{sfxt5:tab:campaign}, for references). 
These data show that the four sources are quite active even outside the outbursts, not
only when observed by XRT, but also by BAT.

\begin{table*}
 \begin{center}
 \caption{BAT on-board detections in the 15--50\,keV band.\label{sfxt5:tab:bat_triggers} }
 \begin{tabular}{rllllrllll}
 \hline
 \noalign{\smallskip}
MJD & Date & Time$^{\mathrm{a}}$ & BAT Trigger N.$^{\mathrm{b}}$ &S/N$^{\mathrm{c}}$  & \hspace{1.2truecm}MJD & 	Date & Time$^{\mathrm{a}}$ & BAT Trigger N.$^{\mathrm{b}}$ &S/N$^{\mathrm{c}}$ \\
  \noalign{\smallskip}
 \hline
 \noalign{\smallskip}							        	  
 & &IGR~J16479$-$4514 & & & & &XTE~J1739$-$302 \\
\noalign{\smallskip\hrule\smallskip}
53612	&2005-08-30	&04:03:28--04:13:52	&152652 (NFI)	  &7.71&	53581	& 2005-07-30	  &00:23:12--00:28:56	   & & \\
53811	&2006-03-17	&08:03:51	        &                 &    &	53663	& 2005-10-20	  &12:53:04	   &	        & \\
53875	&2006-05-20	&17:32:39--17:35:51	&210886 (no slew) &5.78&	53765	& 2006-01-30	  &18:33:43--20:26:15	   &    & \\
53898	&2006-06-12	&06:58:31	        &                 &    &	53798	& 2006-03-04	  &04:52:31	   &	        & \\
53910	&2006-06-24	&20:19:59	        &215914 (no slew) &5.34&	53806	& 2006-03-12	  &07:13:27	   &	        & \\
54095	&2006-12-26	&22:39:43--22:45:03	&                 &    & 	54140	& 2007-02-09	  &17:33:03--17:34:07	   &    & \\
54167	&2007-03-08	&06:04:55	        &                 &    & 	54161	& 2007-03-02	  &13:39:11--15:05:27	   & $^{\mathrm{d}}$   & \\
54196	&2007-04-06	&15:22:55	        &                 &    & 	54168	& 2007-03-09	  &11:09:51--11:14:47	   & $^{\mathrm{d}}$   & \\
54239	&2007-05-19	&19:53:59--20:04:39	&                 &    & 	54269	& 2007-06-18	  &03:09:43--03:10:47	   &282535  (no slew) &6.53 \\
54310	&2007-07-29	&12:07:35	        &286412 (no slew) &9.98&	54411	& 2007-11-07	  &04:38:15--04:42:31	   &    & \\
54320	&2007-08-08	&21:13:51	        &                 &    & 	54564	& 2008-04-08	  &16:43:19--21:28:15	   &308797 (NFI)&7.83 \\
54368	&2007-09-25	&18:14:31	        &                 &    & 	54565	& 2008-04-09	  &00:45:11--00:51:35	   &    & \\
54506	&2008-02-10	&05:35:43	        &                 &    & 	54632	& 2008-06-15	  &06:56:39	   &	       & \\
54535	&2008-03-10	&13:11:03--13:20:39	&                 &    & 	54673	& 2008-07-26	  &14:13:27	   &	       & \\
54544	&2008-03-19	&22:44:47--22:59:59	&306829 (NFI)     &12.02$^{\mathrm{e}}$& 54691 & 2008-08-13 & 23:49:19--23:51:27  &319963 (NFI)&9.15 \\
54572	&2008-04-16	&17:07:11	        &                 &    & 	54692	& 2008-08-14	  &00:04:23--03:06:39	   &319964 (NFI)&11.14 \\
54607	&2008-05-21	&06:03:43--15:38:23	&312068 (no slew) &7.21&	54724	& 2008-09-15	  &12:59:59--13:06:23	   &    & \\
54664	&2008-07-17	&18:10:15	        &                 &    & 	54900	& 2009-03-10	  &18:18:38--18:39:58	   &346069 (NFI)&6.81 \\
54679	&2008-08-01	&03:38:31--03:49:03	&                 &    & 	    &     	  &     	  	   &    & \\
54682	&2008-08-04	&04:06:55--04:14:39	&                 &    & 	    &     	  &     	  	   &    & \\
54687	&2008-08-09	&01:11:51	        &                 &    & 	    &     	  &     	  	   &    & \\
54826	&2008-12-26	&15:55:11--15:57:19	&                 &    & 	    &     	  &     	  	   &    & \\
54860	&2009-01-29	&06:32:06--06:48:14	&341452 (NFI)	  &10.68&	    &     	  &     	  	   &    & \\
\noalign{\smallskip\hrule\smallskip}							        	  
 & &IGR~J17544$-$2619 & & & & & AX~J1841.0$-$0536 \\
\noalign{\smallskip\hrule\smallskip}
53996	&2006-09-18	&09:07:43	        &    && 	 53821   & 2006-03-27	   &00:51:03--08:39:27     &202892 (no slew) & 5.45 \\
53998	&2006-09-20	&11:11:43	        &$^{\mathrm{f}}$    && 	 53834   & 2006-04-09	   &16:33:43	    	   &    &  \\
54009	&2006-10-01	&10:34:47        	&    && 	 53845   & 2006-04-20	   &14:42:31	    	   &    &  \\
54035	&2006-10-27	&00:33:11	        &    && 	 54026   & 2006-10-18	   &10:50:23	    	   &    &  \\
54372	&2007-09-29	&13:21:51--16:36:55	&    && 	 54053   & 2006-11-14	   &10:21:19	    	   &    &  \\
54387	&2007-10-14	&00:39:35--19:57:27	&    && 	 54139   & 2007-02-08	   &22:21:59	    	   &    &  \\
54412	&2007-11-08	&01:31:03--06:16:47	&    && 	 54195   & 2007-04-05	   &12:00:31	    	   &    &  \\
54556	&2008-03-31	&20:50:47	        &308224 (NFI)& 9.10 & 	     &     	   &     		   &    &  \\
54565	&2008-04-09	&18:17:19	        &    && 	     &     	   &     		   &    &  \\
54611	&2008-05-25	&14:43:51--14:55:03	&    && 	     &     	   &     		   &    & \\

  \noalign{\smallskip}
  \hline
  \end{tabular}
  \end{center}
  \begin{list}{}{}
  \item[$^{\mathrm{a}}$]{Time of the start of the BAT trigger, or the time range when on-board detections were obtained. } 
  \item[$^{\mathrm{b}}$]{BAT regular trigger, as was disseminated through GCNs. NFI indicates that there are data from the narrow-field instrument;
 no slew, indicates that \sw\ did not slew to the target.} 
  \item[$^{\mathrm{c}}$]{On-board image significance in units of $\sigma$.}
  \item[$^{\mathrm{d}}$]{Also reported by \citet{Blay2008}. }
  \item[$^{\mathrm{e}}$]{Trigger 306830 had S/N$=21.64$,  see \citet{Romano2008:sfxts_paperII}. }
  \item[$^{\mathrm{f}}$]{Also reported by \citet{Ducci2008}. }
  \end{list}  
\end{table*}

       \subsection{Searching for orbital periodicities in BAT data}

\begin{figure}
\begin{center}
\centerline{\includegraphics[width=9cm,angle=0]{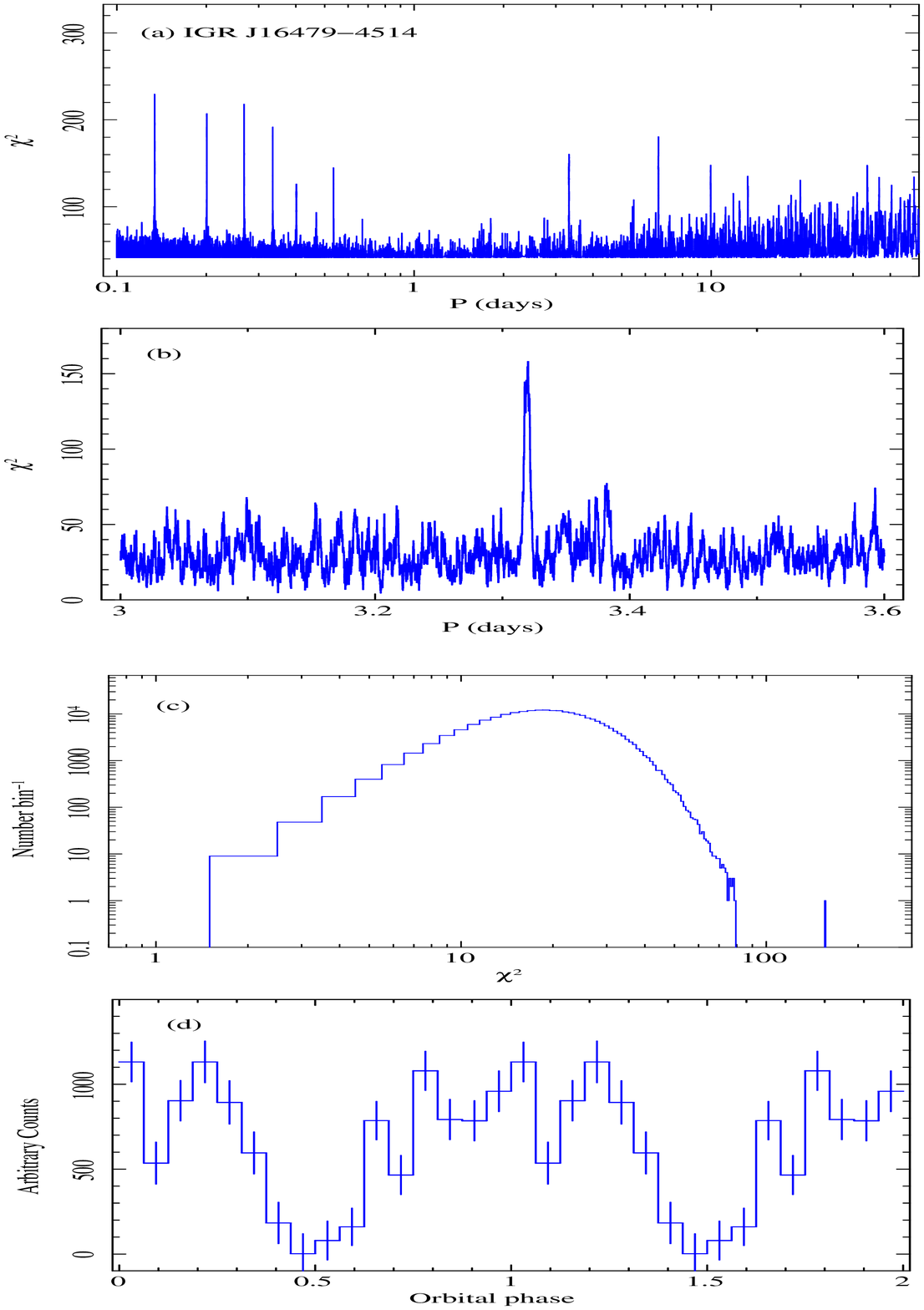}}
\caption{ {\bf (a):} Periodogram of \sw/BAT (15--50\,keV) data for IGR~J16479--4514
                  (MJD range 53413--54829) from the BAT Transient Monitor in the whole time range examined.
                  {\bf (b):} Close-up around the $P=3.32$\,day periodicity. 
                 {\bf (c):} Distribution of $\chi^2$ values.
                 {\bf (d):} Light curve folded at a period $P=3.32$\,day, with 16 phase bins. 
             }
 	\label{sfxt5:fig:bat_periodo_16479}   
 \end{center}
       \end{figure}

\begin{figure}
\begin{center}
\vspace{-2truecm}
\centerline{\includegraphics[width=9cm,angle=0]{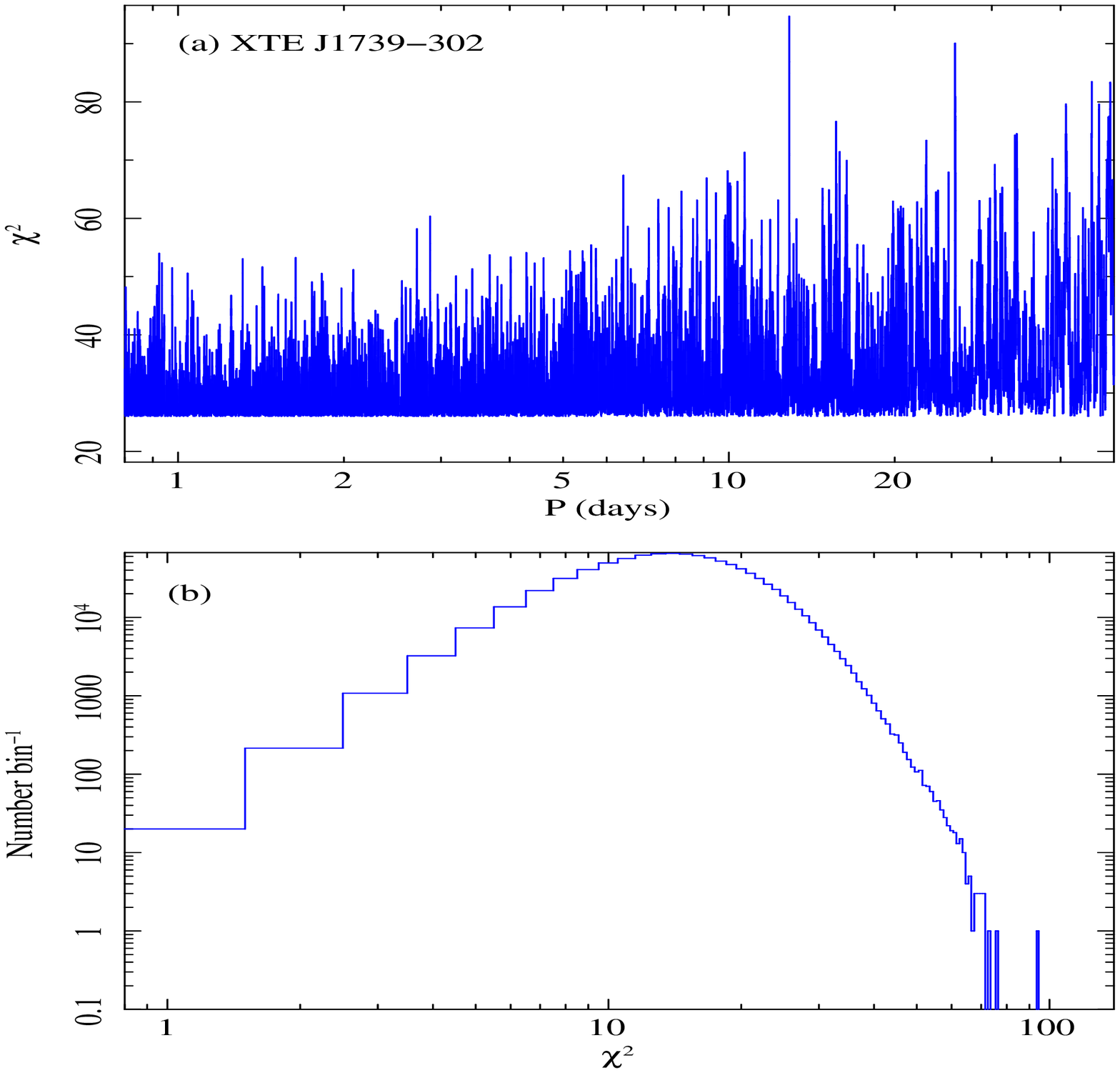}}
\vspace{-2.5truecm}
\caption{ {\bf (a):} Periodogram of \sw/BAT (15--50\,keV) data for XTE~J1739--302
                  (MJD range 53413--54829) from the BAT Transient Monitor in the whole time range examined.
                 {\bf (b):} Distribution of $\chi^2$ values.    }
		\label{sfxt5:fig:bat_periodo_17391}   
       \end{center}
       \end{figure}

We looked for evidence of orbital periodicities in the BAT data. 
We first considered the BAT data for the four sources 
binned orbit-by-orbit in the time range MJD 53413--54829. 
These data were further screened to exclude bad quality points 
(quality flag 1 and 2) and referred to the solar system barycenter (SSB)
by using the task {\sc earth2sun}.

A folding technique was applied to the baricentered
arrival times by searching in the 0.1--50\,d period range and by
building 16-bin pulse profiles with a step given by $P^{2}/(N \,\Delta T)$,
where $N$ is the number of phase bins, and $\Delta T$ is the data span length. 
We find significant evidence for orbital modulation for IGR~J16479--4514
with a best period of 286792$\pm42$\,s  ($P_{\rm orb}=3.3193\pm0.0005$\,days) 
referred to the epoch time MJD 54170.20500213, with a $\chi^2$ value of 155.8. 
As shown in the periodogram in Fig.~\ref{sfxt5:fig:bat_periodo_16479}a,  
this periodicity stands out from the noise and is certainly not due to the 
satellite orbital period or its multiples, as instead is the case for the 
peaks appearing below 1\,day. Peaks at periods higher than 5 days are 
multiples of the  $P_{\rm orb}$. 
A zoomed-in region of the periodogram around the candidate $P_{\rm orb}$ 
is shown in  Fig.~\ref{sfxt5:fig:bat_periodo_16479}b. 
The statistics of the data is not Gaussian, so assessment of the significance of this
periodicity needs to be performed on the noise distribution of $\chi^2$ in the periodogram. 
Fig.~\ref{sfxt5:fig:bat_periodo_16479}c represents 
the noise distribution of the powers of the periodogram 
after removal of the satellite orbital data period (and its multiples), 
and of the multiples of $P_{\rm orb}$. 
In order to evaluate the significance of the signal at $P_{\rm orb}$, 
we fit the distribution for $25< \chi^2 < 80$  with an exponential function
and evaluated the integral of the best-fit function beyond the value of the $\chi^2$ 
obtained at $P_{\rm orb}$. 
This integral yields a number of chance occurencies due to noise of 
$1.24\times10^{-5}$, corresponding to 4.5 standard deviations in Gaussian 
statistics.   
In Fig.~\ref{sfxt5:fig:bat_periodo_16479}d  we show the pulsed profile folded at  $P_{\rm orb}$.
There is clear evidence for an eclipse phase, whose epoch centroid, evaluated by fitting 
the data around the dip with a Gaussian function, is MJD 54171.11$\pm0.05$. 
This confirms the results of \citet{Jain2009:16479}.

Adopting the same techniques for XTE~J1739--302 (Fig.~\ref{sfxt5:fig:bat_periodo_17391}a) 
we find marginal evidence for a signal above the noise at $1111605.1\pm631.3$\,s  
($P_{\rm orb}=12.8658\pm0.0073$\,days)  with a $\chi^2$ value of 94.7. The second 
highest peak in Fig.~\ref{sfxt5:fig:bat_periodo_17391}a is the first multiple of 
$P_{\rm orb}$. The chance probability to obtain this signal at $P_{\rm orb}$ is $3.1\times 10^{-2}$,
corresponding to 2.1 standard deviations in Gaussian statistics.  
By repeating this kind of analysis on 1/4, 1/2, 3/4 and the whole BAT data sample, 
we verified that the power at this $P_{\rm orb}$ increases with time baseline, 
as is expected of signal, as opposed to noise. This strengthens the possibility
that this is a true periodicity. 

No significant evidence for orbital periodicity was found for either IGR~J17544$-$2619 or AX~J1841.0$-$0536.

       \subsection{Searching for spin periodicities in XRT data}

We also looked for evidence of spin periodicities in XRT data. 
For each source, we performed a timing analysis to search for coherent pulsations within
{\it each single observation} with a slow Fourier analysis on the fundamental 
harmonics in the 0.0047--0.199418\,Hz frequency range 
(the latter being the Nyquist frequency of the data set, 
corresponding to a period of 5.01460 s), with the frequency resolution 
$df = 1/(2\Delta T)$ Hz, where $\Delta T$ is the
 length of each observation. 
As the expected power of the pulsed emission is $P_{i} = K \times F_{\rm P}^2 \times N_t+2$, 
we only used observations with a minimum statistic content $N_t>300$ counts, 
that would yield a detection with a significance greater than 3 standard deviations 
for a signal with a pulsed fraction of $F_{\rm P} =0.2$, and $K=0.5$ (sinusoidal profile). 
No significant deviations from a statistically flat 
distribution was revealed in the Fourier spectra of these observations.

In order to reveal the presence of a pulsed signal that could be undetectable in single
observations because of the poor statistics, we
performed a {\it stacked timing analysis} on a larger set of observations. 
This analysis consists in summing the power spectra obtained from single observations 
with a common frequency range (0.005--0.2\,Hz) and resolution $1/2\Delta T_{\rm max}$, 
where $\Delta T_{\rm max}$ is the elapsed time of the longest 
observation. The averaged power spectrum will have a distribution with mean 2 and standard
deviation $2/\sqrt{N}$, where $N$ is the number of summed power spectra. 
However, we cannot add arbitrarily long amounts of data without taking into account
the Doppler modulation due to orbital motion which could destroy the coherence of the 
pulsed signal. 
For IGR~J16479--4514, for which a firm detection of a $P_{\rm orb}$ was obtained, 
we could minimize the effect of the orbital Doppler modulation, 
by summing the spectra which are close in orbital phase. 
Under the simplifying assumption of a circular orbit, we evaluated the orbital phase for each XRT observation
and divided the sample in 4 phase intervals: 0.85--0.15, 0.15--0.35, 0.35--0.65, 0.65--0.85. 
The different amplitudes were chosen to take into account the different values of the tangential 
velocity of the compact object along its orbit.   
In the four stacked spectra we found no evidence for a significant excess above the noise distribution.

       \subsection{Searching for eclipses  in IGR~J16479$-$4514 XRT data}

\begin{figure}
\begin{center}
\centerline{\includegraphics[width=3.5cm,angle=-90]{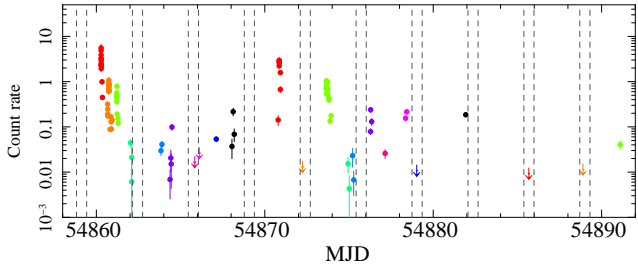}}
\caption{XRT light curve of the \sw/XRT observations after the 2009 January 29 outburst of 
        IGR~J16479$-$4514 in the 0.3--10\,keV band, 
		background-subtracted and corrected for pile-up, PSF losses, and vignetting. 
                Different colours denote different observations. 
                Filled circles are full detections 
                ($S/N > 7$), while downward-pointing arrows are 3-$\sigma$ upper limits.
                  The vertical lines mark the predicted positions of the eclipse.    }
		\label{sfxt5:fig:xrt_lcv_16479_2009}   
       \end{center}
       \end{figure}

The data of the whole XRT campaign on IGR~J16479$-$4514 were sought for the presence of 
eclipses, suggested by \citet{Bozzo2008:eclipse16479} 
on the basis of the analysis of an XMM-Newton observation. 
We created event lists for the whole campaign and selected those inside and outside 
the eclipses, where by `inside the eclipse' we consider the time interval between 
the start of the eclipse as defined by \citet{Bozzo2008:eclipse16479} and 0.6\,d later 
[using the ephemeris from \citep{Jain2009:16479}].
We then calculated the net (subtracted for scaled background) count rate in the two cases. 
We obtain $(6\pm1)\times10^{-3}$ counts s$^{-1}$ (inside) and $0.169\pm0.002$ counts s$^{-1}$ (outside).
Consistent values [$(6\pm3)\times10^{-3}$ and $0.203\pm0.003$ counts s$^{-1}$, respectively]
are measured during the 2009 January outburst. 
This indicates that the source is in two distinct flux levels inside and outside the 
predicted times of the eclipses at the $\sim 50\sigma$ level. 
We also calculated the count rates within individual time slices 
inside eclipses and find that they never exceed 0.013 counts s$^{-1}$. 
We can thus conclude that the XRT data are consistent with the presence of an eclipse 
on the longest baseline so far examined. 
In particular, Fig.~\ref{sfxt5:fig:xrt_lcv_16479_2009} shows the light curve of 
IGR~J16479$-$4514 during the 2009 January 29 outburst with vertical lines marking 
the predicted positions of the eclipse times.

       \subsection{UVOT light curves}

\begin{figure}
\begin{center}
\vspace{-2truecm}
\centerline{\includegraphics[width=9cm,angle=0]{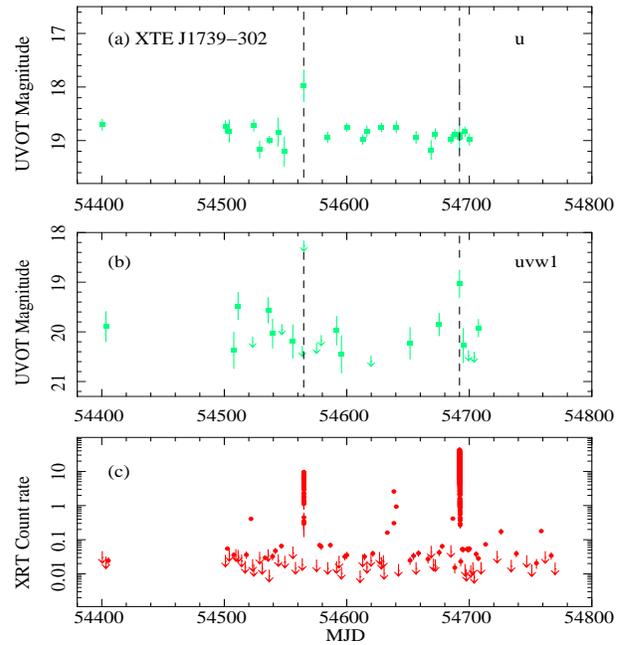}}
\vspace{-2.5truecm}
\caption{Light curves of  XTE~J1739--302. 
         {\bf (a)} \sw/UVOT $u$ light curve;   {\bf (b)}  \sw/UVOT $uvw1$ light curve; 
                {\bf (c)} \sw/XRT light curve. 
		The vertical dashed lines mark the BAT outbursts. 
   }
	\label{sfxt5:fig:uvot_lcv_17391}   
       \end{center}
       \end{figure}

\begin{figure}
\begin{center}
\vspace{-2truecm}
\centerline{\includegraphics[width=9cm,angle=0]{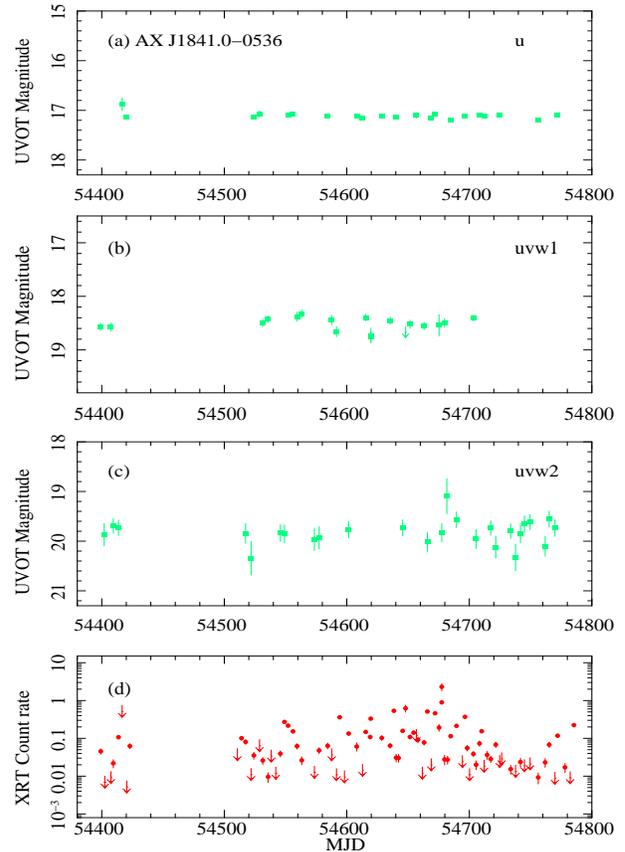}}
\caption{Light curves of  AX~J1841.0$-$0536. 
         {\bf (a)} \sw/UVOT $u$ magnitudes;   {\bf (b)}  \sw/UVOT $uvw1$ magnitudes; 
                {\bf (c)} \sw/XRT $uvw2$ magnitudes. 
               {\bf (d)} \sw/XRT light curve.   }
	\label{sfxt5:fig:uvot_lcv_18410}   
       \end{center}
       \end{figure}

We report for the first time on optical/UV observations performed with UVOT 
simultaneously to our \sw/XRT monitoring of the SFXTs. 
In Fig.~\ref{sfxt5:fig:uvot_lcv_17391}a and \ref{sfxt5:fig:uvot_lcv_17391}b 
we show the UVOT $u$ and $uvw1$ light curves of XTE~J1739$-$302 of the whole campaign. 
The dashed vertical lines mark the two X-ray 
outbursts (2008 April 8, 2008 August 13; Fig.~\ref{sfxt5:fig:uvot_lcv_17391}b). 
The ultraviolet filters only registered upper limits for this source 
during the outbursts.
We consider the correspondence between the X-ray peak and the $u$ magnitude. 
During the first outburst the $u$ band shows an increase of
$\sim 0.9$ mag with respect to the campaign mean or a factor of $\sim2.5$ in flux. 
Depending on the choice of background regions this is a 2--3 $\sigma$ effect.
During the second outburst the $u$ magnitude is consistent with the 
mean for the whole campaign; the $b$ and $v$ magnitudes, collected as part of the 
GRB-chasing filter scheme,  show the same level as in the first one. 
The $uvw1$ magnitudes show a larger degree of variability when compared with the 
$u$ band. 
We also investigated intra-day variability during the outbursts,
but found no significant variation within the errors in all bands.

In Fig.~\ref{sfxt5:fig:uvot_lcv_18410}a,b,c we show the UVOT $u$ and $uvw1$, 
and $uvw2$ light curves of AX~J1841.0$-$0536.  
The $u$ and $uvw1$ are remarkably stable, while the $uvw2$ show some degree of 
variability. We note that the highest point in the $uvw2$ light curve is not
simultaneous with the X--ray peak.

\section{X-ray spectroscopy\label{sfxt5:spectroscopy}}

\begin{table*}
 \begin{center}
 \caption{XRT spectroscopy of the four SFXTs  (2007$+$2008 data set).\label{sfxt5:tab:specfits} }
 \begin{tabular}{crrcccccclc}
 \hline
 \noalign{\smallskip}
Name  & Spectrum & Rate &$N_{\rm H}$& parameter  &Hardness$^{\mathrm{a}}$
            &Flux$^{\mathrm{b}}$   & Luminosity$^{\mathrm{c}}$ &$\chi^{2}_{\rm red}$/dof$^{\mathrm{d}}$  \\ 
Absorbed power law  & & (counts s$^{-1}$)& ($10^{22}$~cm$^{-2}$)    & $\Gamma$  & Ratio  &(2--10 keV)  & (2--10 keV)  &Cstat(\%)  \\
  \noalign{\smallskip}
 \hline
 \noalign{\smallskip}
  IGR~J16479$-$4514  & high     & $>$0.52      &$7.0_{-0.7}^{+0.8}$ &$1.2_{-0.2}^{+0.2}$ &      &120 &5 &$1.1/131$	      \cr
                     & medium   & [0.25--0.52[ &$9.3_{-1.0}^{+1.1}$ &$1.5_{-0.2}^{+0.2}$ &      &54  &3 &$1.0/132$	      \cr
                     & low      & [0.06--0.25[ &$6.7_{-0.7}^{+0.7}$ &$1.4_{-0.2}^{+0.2}$ &      &19  &0.8  &$0.9/144$	      \cr
                     & very low$^{\mathrm{e}}$ & $<$0.06 &$3.3_{+0.0}^{+1.4}$  &$1.5_{-0.4}^{+0.5}$ &      &1.8  &0.04   &482.3(89.3)	      \cr
                     & very low$^{\mathrm{f}}$ &$<$0.06 &$0.3_{-0.3}^{+0.6}$ &$0.3_{-0.4}^{+0.5}$ &$0.9\pm0.3$ &1.8  &0.05   &471.5(50.2)    \cr
\noalign{\smallskip\hrule\smallskip}
     XTE~J1739$-$302 & high     &$>$0.33      &$2.7_{-0.4}^{+0.5}$ &$0.9_{-0.2}^{+0.2}$  &       &120 &1       &$1.1/77$	       \cr
                     & medium   &[0.07--0.33[ &$3.6_{-0.5}^{+0.6}$ &$1.6_{-0.2}^{+0.2}$  &       &15  &0.2   &$0.9/73$		\cr
                     & very low$^{\mathrm{e}}$ &$<$0.07 &$1.7_{-0.0}^{+0.2}$  &$1.3_{-0.3}^{+0.3}$  &    	&0.5  &0.005   &614.3(96.3)	       \cr
                     & very low$^{\mathrm{f}}$ &$<$0.07 &$0.3_{-0.2}^{+0.3}$ &$0.5_{-0.3}^{+0.3}$ &$0.6\pm0.3$ &0.6  &0.006    &598.9(64.7)     \cr
\noalign{\smallskip\hrule\smallskip}							      
   IGR~J17544$-$2619 & high     &$>$0.33       &$1.7_{-0.3}^{+0.3}$ &$1.4_{-0.2}^{+0.2}$ &      &62 &1  &$0.9/75$	         \cr
                     & medium   &[0.07--0.33[  &$2.1_{-0.3}^{+0.3}$ &$1.8_{-0.2}^{+0.2}$ &      &14 &0.3   &$1.0/72$ 	 \cr
                     & very low$^{\mathrm{e}}$ &$<$0.07 &$1.1_{-0.0}^{+0.1}$  &$2.2_{-0.4}^{+0.3}$ &      &0.2 &0.002   &381.8(83.0)	 \cr
                     & very low$^{\mathrm{f}}$ &$<$0.07 &$0.4_{-0.3}^{+0.3}$ &$1.4_{-0.4}^{+0.5}$ &$0.2\pm0.3$  &0.2 &0.003   &372.0(53.1)    \cr
\noalign{\smallskip\hrule\smallskip}							        	  
   AX~J1841.0$-$0536 & high     & $>$0.4       &$2.5_{-0.3}^{+0.3}$ &$1.1_{-0.1}^{+0.1}$ &      &80 &3   &$1.2/110$	         \cr
                     & medium   & [0.18--0.4[  &$3.5_{-0.5}^{+0.5}$ &$1.3_{-0.2}^{+0.2}$ &      &34 &1   &$1.1/102$	         \cr
                     & low      & [0.05--0.18[ &$3.5_{-0.5}^{+0.5}$ &$1.5_{-0.2}^{+0.2}$ &      &11 &0.4   &$1.2/104$	 \cr
                     & very low$^{\mathrm{e}}$ & $<$0.05  &$0.3_{-0.3}^{+0.3}$ &$0.6_{-0.4}^{+0.4}$ &$1.3\pm1.0$ &0.6 &0.02   &449.5(54.7)	 \cr
\noalign{\smallskip\hrule\smallskip}							        	  
Absorbed blackbody   &          & &     &$kT_{\rm bb}$ & Radius (km) & & & & \cr
\noalign{\smallskip\hrule\smallskip}
   IGR~J16479$-$4514 & high     & $>$0.52      &$4.2_{-0.5}^{+0.5}$ &$1.9_{-0.1}^{+0.1}$  &$0.54 ^{+0.06}_{-0.05}$   &110 &4   &$1.2/131$   \cr
                     & medium   & [0.25--0.52[ &$5.7_{-0.6}^{+0.7}$ &$1.8_{-0.1}^{+0.1}$  &$0.43 ^{+0.05}_{-0.04}$   &51 &2   &$1.1/132$    \cr
                     & low      & [0.06--0.25[  &$3.8_{-0.4}^{+0.5}$ &$1.8_{-0.1}^{+0.1}$  &$0.25 ^{+0.03}_{-0.02}$   &17 &0.6   &$1.0/144$	 \cr
                     & very low$^{\mathrm{e}}$ & $<$0.06 &$3.3_{-0.0}^{+0.5}$ &$1.2_{-0.2}^{+0.2}$   &$0.14 ^{+0.05}_{-0.03}$   &1.3 &0.04 & 486.8(99.5)   \cr
                     & very low$^{\mathrm{f}}$ & $<$0.06 &$0.04_{-0.04}^{+0.37}$ &$1.9_{-0.4}^{+0.5}$  &$0.054^{+0.022}_{-0.001}$ &1.4 &0.04 & 462.4(67.3) \cr 
\noalign{\smallskip\hrule\smallskip}
     XTE~J1739$-$302 & high     &$>$0.33      &$1.3_{-0.2}^{+0.3}$ &$1.9_{-0.1}^{+0.2}$  &$0.28^{+0.04}_{-0.03}$    &100 &1   &$1.3/77$	     \cr
                     & medium   &[0.07--0.33[ &$1.5_{-0.3}^{+0.3}$ &$1.5_{-0.1}^{+0.1}$  &$0.16^{+0.02}_{-0.02}$    &13 &0.1	&$0.8/73$	     \cr
                     &very low$^{\mathrm{e}}$ &$<$0.07  &$1.7_{-0.0}^{+0.1}$ &$1.3_{-0.1}^{+0.2}$  &$0.04^{+0.01}_{-0.01}$    &0.5 &0.004   &646.1(100.0)	     \cr
                     &very low$^{\mathrm{f}}$  &$<$0.07 &$0.0_{-0.0}^{+0.1}$ &$1.7_{-0.2}^{+0.3}$ &$0.022^{+0.005}_{-0.004}$ &0.5 &0.004  & 597.8(49.8)   \cr
\noalign{\smallskip\hrule\smallskip}							        	  
   IGR~J17544$-$2619 & high     &$>$0.33      &$0.6_{-0.1}^{+0.1}$ &$1.4_{-0.1}^{+0.1}$  & $0.44^{+0.05}_{-0.04}$ &53 &0.9   &$1.3/75$		   \cr
                     & medium   &[0.07--0.33[ &$0.8_{-0.2}^{+0.2}$ &$1.2_{-0.1}^{+0.1}$  & $0.30^{+0.04}_{-0.03}$ &12 &0.2   &$1.1/72$		    \cr
                     & very low$^{\mathrm{e}}$ &$<$0.07 &$1.1_{-0.0}^{+0.1}$ &$0.8_{-0.1}^{+0.1}$ & $0.08^{+0.03}_{-0.02}$ &0.2 &0.002  &425.9(99.9)	   \cr
                     & very low$^{\mathrm{f}}$ &$<$0.07 &$0.0_{-0.0}^{+0.1}$ &$1.1_{-0.1}^{+0.2}$ &$0.04^{+0.01}_{-0.01}$ &0.2 &0.002   & 381.8(50.5) \cr
\noalign{\smallskip\hrule\smallskip}							        	  
   AX~J1841.0$-$0536 & high     & $>$0.4       &$1.2_{-0.2}^{+0.2}$ &$1.7_{-0.1}^{+0.1}$ & $0.51^{+0.05}_{-0.05}$ &71	 &2   &$1.5/110$       \cr
                     & medium   & [0.18--0.4[  &$1.7_{-0.3}^{+0.3}$ &$1.6_{-0.1}^{+0.1}$ & $0.39^{+0.03}_{-0.03}$ &30	 &1   &$1.2/102$	  \cr
                     & low      & [0.05--0.18[ &$1.5_{-0.2}^{+0.3}$ &$1.5_{-0.1}^{+0.1}$ & $0.24^{+0.03}_{-0.02}$ &9.9	 &0.3   &$1.0/104$	    \cr
                     & very low$^{\mathrm{e}}$ & $<$0.05  &$0.0_{-0.0}^{+0.1}$ &$1.7_{-0.3}^{+0.4}$  &$0.04^{+0.01}_{-0.01}$ &0.5 &0.02   & 451.7(50.0) \cr   
  \noalign{\smallskip}
  \hline
  \end{tabular}
  \end{center}
  \begin{list}{}{}
  \item[$^{\mathrm{a}}$]{Hardness ratio 4--10 keV / 0.2--4 keV .  }	
  \item[$^{\mathrm{b}}$]{Average observed 2--10\,keV fluxes in units of 10$^{-12}$~erg~cm$^{-2}$~s$^{-1}$.}
  \item[$^{\mathrm{c}}$]{Average 2--10\,keV X--ray luminosities in units of 10$^{35}$~erg~s$^{-1}$ calculated 
         adopting distances determined by \citet{Rahoui2008}. }
  \item[$^{\mathrm{d}}$]{Reduced $\chi^{2}$ and dof, or Cash statistics Cstat and percentage of realizations  ($10^4$ trials) with statistic $>$ Cstat. }
  \item[$^{\mathrm{e}}$]{Fit performed with constrained column density (see Sect~\ref{sfxt5:spec_out_of_outburst}). }
  \item[$^{\mathrm{f}}$]{Fit performed with free column density  (see Sect~\ref{sfxt5:spec_out_of_outburst}). }
  \end{list}  
\end{table*}

\begin{figure*}
\begin{center}
\vspace{-3truecm}
\centerline{\includegraphics[width=18cm,height=18cm,angle=0]{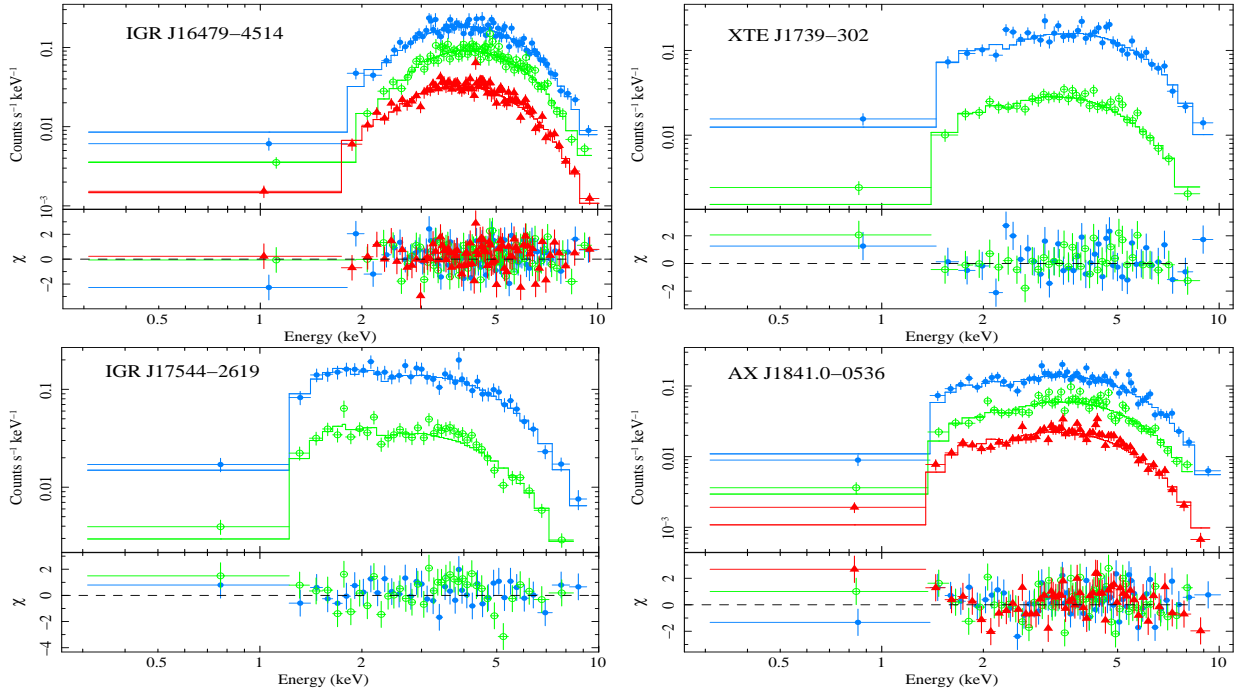}}
\vspace{-5truecm}
\caption[XRT spectroscopy, spectra and residuals]{Spectroscopy of the 2007--2008 
                  observing campaign.
            Upper panels: XRT/PC data fit with an absorbed power law. 
            Lower panels: the residuals of the fit (in units of standard deviations). 
            Filled blue circles, green empty circles, and red filled triangles
            mark high, medium, and low states, respectively. 
                 }
	\label{sfxt5:fig:specfits}   
       \end{center}
       \end{figure*}

\subsection{The 2009 January 29 outburst of IGR~J16479$-$4514}

In response to the 2009 January 29 outburst of IGR~J16479$-$4514,
\sw\  performed a delayed slew, so that the NFI were on target $>800$\,s
after the trigger \citep{Romano2009:atel1920}. At this point, the flux registered by the BAT was rather low,
and meaningful broad-band (XRT$+$BAT) spectroscopy is not possible, given the 
$107$\,s overlap in the observations. 
Therefore, here we only report the fits to the XRT data. 
An absorbed power-law model yielded an absorbing column of 
$N_{\rm H}=(7.1_{-2.0}^{+2.6})\times 10^{22}$ cm$^{-2}$, 
a photon index $\Gamma=1.63_{-0.46}^{+0.53}$,  
and an unabsorbed flux in the 2--10\,keV band is $2\times10^{-10}$ erg cm$^{-2}$ s$^{-1}$. 
The X-ray spectrum extracted from segment 00030296087, when the source showed rebrightening
\citep{LaParola2009:atel1929}, was also fit with 
an absorbed power-law model, obtaining $\Gamma=1.29_{-0.46}^{+0.53}$, 
$N_{\rm H}=(7.1_{-2.0}^{+2.6})\times 10^{22}$ cm$^{-2}$,  
and unabsorbed flux in the 2--10\,keV band of $3\times10^{-10}$ erg cm$^{-2}$ s$^{-1}$. 
These results are generally consistent with the ones found in \citet{Romano2008:sfxts_paperII}, which describes the
outburst of this source which occurred on 2008 March 19, 315 days earlier. 
We also note that, as found in \citet{Romano2008:sfxts_paperII},   the derived $N_{\rm H}$ is in excess 
of the one along the line of sight, $1.87\times 10^{22}$ cm$^{-2}$. 
The broad-band spectroscopy of the other outbursts caught during the campaign 
has already been reported on elsewhere 
[\citet{Romano2008:sfxts_paperII,Sidoli2009:sfxts_paperIV,Sidoli2009:sfxts_paperIII} ; 
see Table~\ref{sfxt5:tab:campaign}]      
and we will also summarize them below.

\subsection{Out-of-outburst X-ray spectroscopy\label{sfxt5:spec_out_of_outburst}} 

\begin{figure*}
\begin{center}
\vspace{-3truecm}
\centerline{\includegraphics[width=18cm,height=18cm,angle=0]{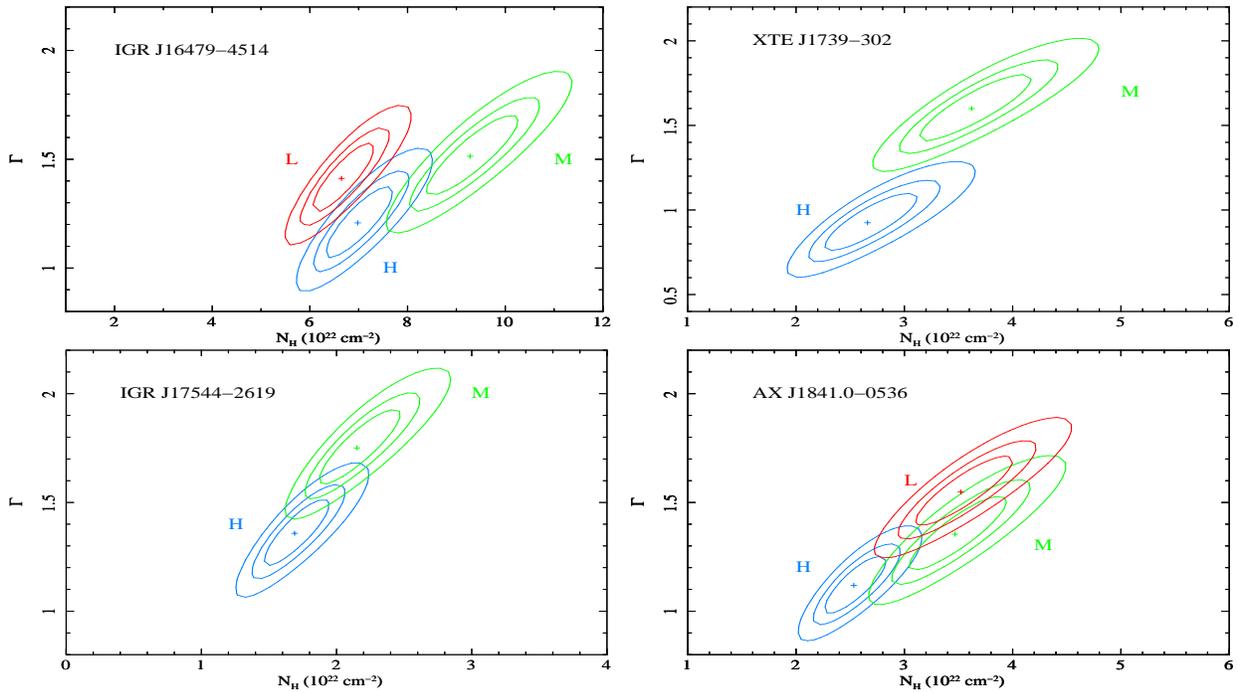}}
\vspace{-5truecm}
\caption[XRT spectroscopy, contour levels]{XRT time selected spectroscopy. 
                  The $\Delta \chi^2 = 2.3$, 4.61 and 9.21 contour levels for the column density in 
            units of $10^{22}$ cm$^{-2}$ vs. the photon index, with best fits indicated by crosses. The color scheme is the
            same as in Fig.~\ref{sfxt5:fig:specfits}. The labels L, M, and H mark low, medium, and high states, respectively.  
                 }
	\label{sfxt5:fig:specfits2}   
       \end{center}
       \end{figure*}

In the remainder of this section we concentrate on the out-of-outburst emission. 
To characterize the spectral properties of the sources in several states, 
we accumulated the events in each observation when 
the source was not in outburst and a detection was achieved. 
For these events we estimated from the light curves (binned at a 100\,s resolution) 
three count rate levels, $CR_1$, $CR_2$, and $CR_3$ (reported in Table~\ref{sfxt5:tab:specfits}) 
that would yield comparable statistics (and at least 1600 counts) in the ranges
$CR_1<CR<CR_2$ (low), $CR_2<CR<CR_3$ (medium), and $CR>CR_3$ (high). 
If the statistics did not allow this, then we only considered two intensity levels 
(high and medium). 
Exposure maps were created for each of these intensity-selected event files.  
We then combined the intensity-selected event files (and their exposure maps) 
and extracted a single spectrum for each source by 
integrating over all the available observing time within these 
intensity limits.
Ancillary response files were generated with {\sc xrtmkarf},
and they account for different extraction regions, vignetting, and
PSF corrections.
The spectra were rebinned with a minimum of
20 counts per energy bin to allow $\chi^2$ fitting. 
Each spectrum was fit in the energy range 0.3--10\,keV 
with a single absorbed power law, or an absorbed black body. 

We also extracted spectra from the event list accumulated from all observations  
for which no detections were obtained as single exposures (very low). 
These spectra consisted of $\sim 100$ counts each, so  
Cash \citep{Cash1979} statistics and spectrally unbinned data were used, instead. 
When fitting with free parameters, the best fit value for $N_{\rm H}$ turned out to be 
consistent with 0, i.e., well below the column derived from optical spectroscopy.
In this case, we performed further fits by adopting as lower limit on the absorbing 
column the value derived from the Galactic extinction estimate along the line of 
sight to each source from \citet{Rahoui2008}, with a conversion into 
Hydrogen column, $N_{\rm H} = 1.79 \times 10^{21} A_V $ cm$^{−2}$ \citep{avnh}. 
As a comparison, we also report the simple hardness ratios.  

The spectra and contour plots of photon index vs.\ column density 
are shown in Fig.~\ref{sfxt5:fig:specfits} and \ref{sfxt5:fig:specfits2}, respectively, 
while the spectral parameters are reported in Table~\ref{sfxt5:tab:specfits}, where 
we also report the average 2--10\,keV luminosities 
calculated by adopting distances determined by \citet{Rahoui2008} 
from optical spectroscopy of the supergiant companions (4.9~kpc for IGR~J16479$-$4514, 2.7~kpc 
for XTE~J1739$-$302, and 3.6~kpc for IGR~J17544$-$2619). 
For AX~J1841.0$-$0536 two estimates of the distance are available, 
\citet[][$3.2_{-1.5}^{+2.0}$\,kpc]{Nespoli2008} and \citet[][$6.9\pm1.7$\,kpc]{Sguera2009}, 
and we assumed a distance of 5~kpc, which is consistent with both.

\begin{figure}
\begin{center}
\centerline{\includegraphics[width=9cm,angle=0]{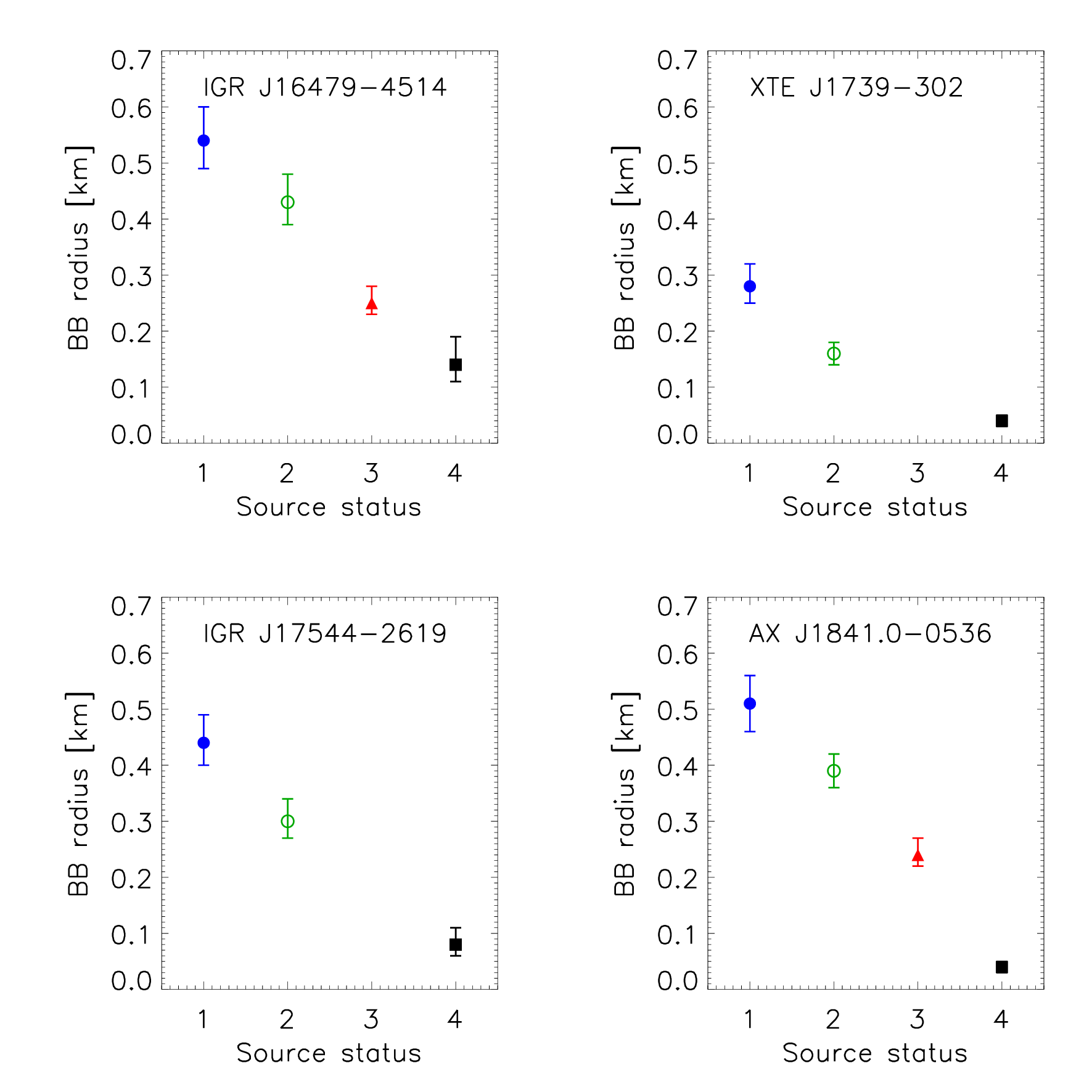}}
\caption[]{Blackbody radii as a function of source state. The color coding is the same as
                  in Fig.~\ref{sfxt5:fig:specfits} and \ref{sfxt5:fig:specfits2}. For the `very low' state,
                  we used the best fits with a constrained $N_{\rm H}$.
                  }
	\label{sfxt5:fig:bbradii}   
       \end{center}
       \end{figure}

We note that spectral fits with an absorbed blackbody always result in blackbody 
radii of a few hundred meters (at the source distances, see Table~\ref{sfxt5:tab:specfits}, 
and Fig.~\ref{sfxt5:fig:bbradii}), 
consistent with being emitted from a small portion of the neutron star 
surface, very likely the neutron star polar caps  \citep{Hickox2004}.  
Indeed, since SFXTs are wind accretors, 
alternative origins for the small emitting region, such as 
small hot regions in an accretion disk, can be discarded.

\section{Discussion}

We  report on the results of an entire year of monitoring campaign with \sw\ of
a sub-sample of SFXTs. For the first time it is possible to investigate in depth 
the long-term properties of this new class of puzzling X--ray transients,
assessing the characteristics of three different source states: the bright outbursts,
the intermediate intensity state, and the quiescence.

During this first year of monitoring, we have obtained multi-wavelength 
observations of 5 outbursts of 3 different sources 
(see Table~\ref{sfxt5:tab:campaign}). 
As reported in \citet{Romano2008:sfxts_paperII,Sidoli2009:sfxts_paperIV,Sidoli2009:sfxts_paperIII}, 
we  studied  the broad-band simultaneous spectra 
(0.3--150\,keV) of three SFXTs. They can be fit with models
traditionally adopted for accreting X--ray pulsars (absorbed 
cutoff power laws), even in the objects where proof of the presence 
of a neutron star (as derived from a spin period) is still unavailable. 
Considerable differences can be found in the
behaviour of the absorbing column among the examined cases,
and the new data from the 2009 January 29 outburst
of IGR~J16479$-$4514 fit well in this picture.

Our \sw\   monitoring campaign has  demonstrated for the first time that 
X--ray emission from SFXTs is still present outside the bright outbursts, 
although at a much lower level (10$^{33}$--10$^{34}$~erg~s$^{-1}$). 
This was already emerging from the first four months of this campaign \citep{Sidoli2008:sfxts_paperI}, 
but now we have accumulated enough statistics to allow intensity 
selected spectroscopy of the out-of-outburst emission. Spectral fits performed  
adopting simple models, such as an absorbed power law or a blackbody 
(more complex models were not required by the data) result in 
hard power law photon indices (always in the range $\Gamma$$\sim$0.8--2) or in 
hot black bodies (kT$_{\rm BB}$$\sim$1--2~keV). It is remarkable that the 
statistics now accumulated  allow us 
to constrain well the spectral parameters of the intermediate level of 
X--ray emission: in particular, when a blackbody model is adopted, the 
resulting radii of the emitter for all 4 SFXTs (and all the intensity 
states) is {\em always only } a few hundred meters (note that even with 
several kpc of uncertainty in the distance determination, the emitting 
regions are always significantly smaller than the neutron star radius). 
This is clearly indicative of an emitting region which is only a fraction of 
the neutron star surface, and can be associated in a natural way with the 
polar caps of the neutron star,  \citep{Hickox2004}. 
This evidence, coupled with the high 
level of flux variability and  hard X--ray spectra, 
strongly supports the 
fact that the intermediate and low intensity level of SFXTs is 
produced by the accretion of matter onto the neutron star, demonstrating 
that SFXTs are sources which do not spend most of their 
lifetime in quiescence. This observational evidence rules out all models for the SFXTs 
emission which predict that the SFXTs are in quiescence when they are not 
in outburst \citep[e.g., ][]{Bozzo2008}. Even the accumulation of matter onto the
neutron star magnetosphere is very difficult to reconcile with our determination
of the emitting radius, which is always smaller than 1~km. 

After following the X-ray light curves of the four SFXTs for one year,
we have obtained the first assessment of how long each source in our sample 
spends in each state using a systematic monitoring. 
The duty-cycle of inactivity is $\sim 17, 28, 39, 55$\,\%  (5\,\% uncertainty), 
for IGR~J16479$-$4514 
AX~J1841.0$-$0536, XTE~J1739--302, and IGR~J17544$-$2619,  
respectively. For IGR~J16479$-$4514 a contribution to the time spent in 
inactivity is due to the X--ray eclipses, hence the above 17\,\% is in fact
an upper limit to the true quiescent time. 
In the latter three SFXTs this inactivity duty cycle, where the sources are 
undetected with \sw, can be associated with the true quiescence (that is, 
no accretion) and/or with an accretion at a very low rate. 
Thus, the quiescence in these transients is a rare state.

The lowest luminosity level  we could monitor 
(`very low' intensity level in Table~\ref{sfxt5:tab:specfits})
with \sw\ is reached in XTE~J1739--302 (6$\times$10$^{32}$~erg~s$^{-1}$, 2--10 keV) and 
in IGR~J17544$-$2619 (3$\times$10$^{32}$~erg~s$^{-1}$).
This latter value is consistent with the quiescent
state observed in IGR~J17544$-$2619 during a $Chandra$ observation 
[5.2$\pm{1.3}$$\times$10$^{32}$~erg~s$^{-1}$, 0.5--10 keV; \citet{zand2005}], although the two spectra
are very different.  During the $Chandra$ observation the spectrum was very soft, likely thermal
(fitted with a power law resulted in a photon index $\Gamma$=5.9$\pm{1.2}$), whereas our accumulated
spectrum during the very low intensity state is much harder ($\Gamma$$\sim$1--2), very likely implying low rate accretion
onto the compact object. 
The lowest level of X--ray emission ever detected from  XTE~J1739--302 has been observed
with ASCA [1.1$\times$10$^{-12}$~erg~cm$^{-2}$~s$^{-1}$, 2--10 keV; \citet{Sakano2002}], which is
much lower than that observed with \sw. This comparison, together with the hard spectrum and the small
emitting radius (see Table~\ref{sfxt5:tab:specfits}) observed with \sw\ implies that we 
have not reached the quiescent state in this source. 
Besides  IGR~J17544$-$2619, the only other SFXTs where the quiescence (characterized by a soft thermal spectrum
and a very low luminosity of $\sim$10$^{32}$~erg~s$^{-1}$) has been caught is IGR~J08408--4503 \citep{Leyder2007}.

The low intensity level we observe with \sw\ in IGR~J16479$-$4514 is consistent with it being due to the 
X--ray eclipses. The possibility of an X--ray eclipse in this transient was originally suggested
by \citet{Bozzo2008:eclipse16479}, based on the variability of the iron line emission
observed during an XMM-Newton observation.  Then, \citet{Jain2009:16479} found a periodicity
at 3.32~days, suggesting it as a possible orbital period. 
We were able to confirm this periodicity from our independent analysis of the BAT data. 

If we assume that this periodicity is of orbital origin, and consider a duration of the 
X$-$ray eclipse of 0.6~days \citep{Jain2009:16479}, then the inclination {\em i} of the system
can be derived from the eclipse semi-angle, $\theta_e$, for an assumed supergiant radius 
[$R_{\rm OB} = 23.8$~R$_{\odot}$ for a O8.5 supergiant, \citet{Vacca1996}] as 
%
$R_{\rm OB} = a \sqrt(\cos^2 i + \sin^2 i \sin^2 \theta_e)$.
%
Assuming the system parameters previously adopted for IGR~J16479--4514,
we obtain a binary separation $a \approx 2 \times 10^{12}$~cm, implying an orbital inclination of 
$i \approx 40^{\circ}$.
  
This 3.32~days periodicity, if interpreted as the orbital period of the binary system, is puzzling,
and is very difficult to reconcile with all the mechanisms proposed to explain the SFXTs phenomenon.
The out-of-eclipse average X$-$ray luminosity of IGR~J16479--4514
is $L_{\rm obs} \approx 10^{34}-10^{35}$~erg~s$^{-1}$. It can be compared
with the X$-$ray emission expected from Bondi-Hoyle accretion onto a neutron star.
Let us assume a circular orbit (very likely, given the short period), 
a stellar mass of $M_{\rm OB} = 30$~M$_{\odot}$, 
a radius $R_{\rm OB} = 23.8$~R$_{\odot}$ for the supergiant \citep{Vacca1996}, 
a beta-law velocity for the supergiant wind $v(r)=v_{\infty}(1- R_{\rm OB}/r)^{\beta}$ 
with $\beta=1$, and a conservative high terminal velocity $v_{\infty}=2000$~km~s$^{-1}$.
Under these assumptions the X$-$ray luminosity produced by the wind accretion 
for a reasonable choice of the wind mass loss rate 
($\dot{M}_{th} \approx 10^{-6}$~M$_{\odot}$~yr$^{-1}$)  
is $\sim$$10^{37}$~erg~s$^{-1}$, 
about 2--3 order of magnitude higher than the observed luminosity.
On the other hand, the observed low luminosity can be obtained only at 
a wind mass loss rate of  $\dot{M}_{\rm obs} \approx 10^{-8} - 10^{-9}$~M$_{\odot}$~yr$^{-1}$,
which is not reasonable for a O8.5 supergiant.

A viable explanation to this inconsistency could be that the 3.32~days periodicity is not
orbital, but is only one of the periodicities predicted in our model for the explanation of
SFXTs outbursts, that is the time interval between the periodically recurrent flares when
the neutron star passes throught the preferential plane for the outflowing wind from the supergiant,
twice per orbit; thus, the true orbital period can be much longer than this periodicity found [see the
different geometries proposed in \citet{Sidoli2007}].

\section*{Acknowledgments}

We would like to thank our many collaborators, who helped along the way during this large project,
and M.\ Colpi, who organized a `Neutron Star Day' in Milano in 2006, during which so many 
new ideas came forth and unexpected collaborations were created... and who basically got this all started. 
Then, we would like to thank A.\ Bazzano, A.\ Cucchiara, S.\ Mereghetti, T.\ Mineo, and P.\ Ubertini 
for helpful discussions. 
We thank the {\it Swift} team duty scientists and science planners  P.J.\ Brown, M.\ Chester,
S.\ Hunsberger,  
J.\ Racusin, and M.C.\ Stroh 
for their dedication and willingness to accomodate our sudden requests
in response to outbursts during this monitoring effort. 
We also thank the remainder of the {\it Swift} XRT and BAT teams,
J.A.\ Nousek and S.\ Barthelmy in particular, for their invaluable help and support with
the planning and execution of the observing strategy. 
This work was supported in Italy by contracts ASI I/088/06/0 and I/023/05/0, at
PSU by NASA contract NAS5-00136. 
HAK was supported by the {\it Swift } project. 
DNB and JAK acknowledge support from NASA contract NAS5-00136. 
PR thanks INAF-IASF Milano and LS INAF-IASF Palermo, 
where some of the work was carried out, 
for their kind hospitality. 
We also thank the anonymous referee for comments that helped improve the paper.

\setcounter{table}{4}
 \begin{table*}
 \begin{center}
 \caption{Observation log for IGR~J16479$-$4514.\label{sfxts:tab:alldata16479}}
 \begin{tabular}{llllll}
 \hline
 \noalign{\smallskip}
 Sequence & Instrument/Mode & Start time (UT) &  End time (UT) &  Net Exposure \\
            &     & (yyyy-mm-dd hh:mm:ss) & (yyyy-mm-dd hh:mm:ss) &  (s) \\ 
  \noalign{\smallskip}
 \hline
 \noalign{\smallskip}
00030296005		&XRT/PC  &2007-10-26 08:08:36	&2007-10-26 09:42:57	&	1176	\\
00030296006		&XRT/PC  &2008-01-19 22:58:28	&2008-01-19 23:13:58	&	927	\\
00030296007		&XRT/PC  &2008-01-22 20:07:57	&2008-01-22 20:25:58	&	1079	\\
00030296008		&XRT/PC  &2008-01-26 18:56:04	&2008-01-26 19:14:57	&	1133	\\
00030296009		&XRT/PC  &2008-01-29 08:03:13	&2008-01-29 08:20:56	&	1062	\\
00030296010		&XRT/PC  &2008-02-05 00:48:59	&2008-02-05 16:58:56	&	3056  \\ 
00030296011	&XRT/PC &	2008-02-09 18:53:39	&	2008-02-09 23:50:58	&	2749	\\
00030296012	&XRT/PC &	2008-02-10 01:16:40	&	2008-02-10 04:36:58	&	873	\\
00030296013	&XRT/PC &	2008-02-15 08:12:38	&	2008-02-15 10:01:25	&	1254	\\
00030296014	&XRT/PC &	2008-02-18 06:49:05	&	2008-02-18 10:12:57	&	804	\\
00030296015	&XRT/PC &	2008-02-20 00:33:38	&	2008-02-20 00:46:58	&	798	\\
00030296016	&XRT/PC &	2008-02-23 21:44:31	&	2008-02-23 23:35:57	&	1667	\\
00030296017	&XRT/PC &	2008-02-26 23:38:06 &	2008-02-26 23:52:56	&	890	\\
00030296018	&XRT/PC    &       2008-03-01 23:38:51     &       2008-03-01 23:59:59     &       1266    \\
00030296019	&XRT/PC    &	   2008-03-03 01:50:45     &	   2008-03-03 05:07:56     &	   1083    \\
00030296020	&XRT/PC    &	   2008-03-06 00:26:49     &	   2008-03-06 02:12:58     &	   1098    \\
00030296021	&XRT/PC    &	   2008-03-10 09:05:42     &	   2008-03-10 12:15:56     &	   1261    \\
00030296022	&XRT/PC    &	   2008-03-11 20:00:41     &	   2008-03-11 23:22:56     &	   2985    \\
00030296023	&XRT/PC    &	   2008-03-12 18:46:44     &	   2008-03-12 23:51:49     &	   2796    \\
00030296024	&XRT/PC    &	   2008-03-13 04:14:40     &	   2008-03-13 04:34:56     &	   1213    \\
00030296025	&XRT/PC    &	   2008-03-15 04:37:49     &	   2008-03-15 04:55:57     &	   1086    \\
00030296026	&XRT/PC    &	   2008-03-16 07:56:54     &	   2008-03-16 08:15:30     &	   1116    \\
00030296027	&XRT/PC    &	   2008-03-17 06:34:08     &	   2008-03-17 08:21:48     &	   1215    \\
00030296028	&XRT/PC    &	   2008-03-18 06:31:54     &	   2008-03-18 16:21:56     &	   2333    \\
00030296029	&XRT/WT    &	   2008-03-19 19:38:51     &	   2008-03-19 22:46:00     &	   67	   \\
00030296029	&XRT/PC    &	   2008-03-19 19:38:56     &	   2008-03-19 19:47:57     &	   542     \\
00306829000	&XRT/WT    &	   2008-03-19 22:46:47     &	   2008-03-19 23:55:26     &	   884     \\
00306829000	&XRT/PC    &	   2008-03-19 23:55:27     &	   2008-03-20 00:00:53     &	   304     \\
00030296030	&XRT/PC    &	   2008-03-21 02:05:29     &	   2008-03-21 03:48:58     &	   835     \\
00030296031	&XRT/PC    &	   2008-03-21 21:22:28     &	   2008-03-21 23:05:57     &	   835     \\
00030296032	&XRT/PC    &	   2008-03-23 05:25:03     &	   2008-03-23 05:35:30     &	   625     \\
00030296033	&XRT/PC    &	   2008-03-24 05:36:51     &	   2008-03-24 06:50:56     &	   1040    \\
00030296034	&XRT/PC    &	   2008-04-10 15:17:37     &	   2008-04-10 20:10:57     &	   905     \\
00030296035	&XRT/PC    &	   2008-05-02 22:12:26     &	   2008-05-02 22:25:56     &	   809     \\
00030296036	&XRT/PC    &	   2008-05-06 20:40:14     &	   2008-05-06 20:43:39     &	   205     \\
00030296037	&XRT/PC    &	   2008-05-10 09:53:10     &	   2008-05-10 10:10:56     &	   1066    \\
00030296038	&XRT/PC    &	   2008-05-14 08:33:12     &	   2008-05-14 08:52:58     &	   1186    \\
00030296039	&XRT/PC    &	   2008-05-22 20:54:24     &	   2008-05-22 22:45:56     &	   2379    \\
00030296040	&XRT/PC    &	   2008-05-26 21:05:40     &	   2008-05-26 21:28:57     &	   1396    \\
00030296041	&XRT/PC    &	   2008-05-30 08:40:01     &	   2008-05-31 18:10:58     &	   995     \\
00030296042	&XRT/PC    &	   2008-06-03 20:15:44     &	   2008-06-03 20:34:57     &	   1153    \\
00030296043	&XRT/PC    &	   2008-06-07 16:05:45     &	   2008-06-07 16:09:56     &	   250     \\
00030296044	&XRT/PC    &	   2008-06-11 08:22:02     &	   2008-06-11 09:57:56     &	   985     \\
00030296046	&XRT/PC    &	   2008-06-19 17:03:51     &	   2008-06-19 18:51:58     &	   1510    \\
00030296047	&XRT/PC    &	   2008-06-23 09:12:34     &	   2008-06-23 09:25:56     &	   802     \\
00030296048	&XRT/PC    &	   2008-06-28 00:35:19     &	   2008-06-28 04:59:57     &	   970     \\
00030296049	&XRT/PC    &	   2008-07-01 05:12:34     &	   2008-07-01 05:28:57     &	   982     \\
00030296050	&XRT/PC    &	   2008-07-05 13:53:11     &	   2008-07-05 14:10:57     &	   1066    \\
00030296052	&XRT/PC    &	   2008-07-13 09:51:01     &	   2008-07-13 11:34:56     &	   948     \\
00030296053	&XRT/PC    &	   2008-07-17 13:26:34     &	   2008-07-17 13:30:44     &	   250     \\
00030296055	&XRT/PC    &	   2008-07-21 02:13:37     &	   2008-07-21 02:29:57     &	   980     \\
00030296056	&XRT/PC    &	   2008-07-25 02:53:34     &	   2008-07-25 03:10:57     &	   1043    \\
00030296057	&XRT/PC    &	   2008-07-29 09:41:29     &	   2008-07-29 09:58:57     &	   1048    \\
00030296058	&XRT/PC    &	   2008-08-02 13:00:42     &	   2008-08-02 14:46:58     &	   933     \\
00030296059	&XRT/PC    &	   2008-08-06 02:09:08     &	   2008-08-06 02:25:56     &	   1007    \\
00030296060	&XRT/PC    &	   2008-08-10 00:57:00     &	   2008-08-10 01:11:57     &	   896     \\
  \end{tabular}
  \end{center}
  \end{table*}

\setcounter{table}{4}
 \begin{table*}
 \begin{center}
 \caption{Observation log for IGR~J16479$-$4514. Continued}
 \begin{tabular}{llllll}
 \hline
 \noalign{\smallskip}
 Sequence & Instrument/Mode & Start time (UT) &  End time (UT) &  Net Exposure \\
            &     & (yyyy-mm-dd hh:mm:ss) & (yyyy-mm-dd hh:mm:ss) &  (s) \\ 
  \noalign{\smallskip}
 \hline
 \noalign{\smallskip}
00030296061	&XRT/PC    &	   2008-08-14 19:10:16     &	   2008-08-14 19:25:56     &	   939     \\
00030296062	&XRT/PC    &	   2008-08-18 11:27:53     &	   2008-08-18 11:43:56     &	   963     \\
00030296063	&XRT/PC    &	   2008-08-22 07:01:50     &	   2008-08-22 23:07:57     &	   755     \\
00030296065	&XRT/PC    &	   2008-08-30 01:29:10     &	   2008-08-30 03:15:56     &	   1236    \\
00030296066	&XRT/PC    &	   2008-09-11 23:18:06     &	   2008-09-11 23:32:56     &	   890     \\
00030296067	&XRT/PC    &	   2008-09-15 07:46:25     &	   2008-09-15 08:02:56     &	   990     \\
00030296070	&XRT/PC    &  	   2008-09-27 05:41:07	&	2008-09-27 05:58:55	&	1068	 \\ 
00030296071     &XRT/PC    &       2008-10-01 15:49:11     &       2008-10-01 16:07:57     &       1126    \\
00030296073     &XRT/PC    &       2008-10-12 23:14:15     &       2008-10-12 23:28:55     &       880     \\
00030296074     &XRT/PC    &       2008-10-17 01:27:07     &       2008-10-17 03:05:57     &       471     \\
00030296075     &XRT/PC    &       2008-10-21 04:46:15     &       2008-10-21 05:00:57     &       883     \\
00030296076     &XRT/PC    &       2008-10-25 22:48:38     &       2008-10-25 23:02:05     &       805     \\
00341452000 & XRT/WT &2009-01-29 06:46:53     &2009-01-29 08:16:49     &46      \\
00341452000 & XRT/PC &2009-01-29 06:47:35     &2009-01-29 08:39:04     &2628	\\
00030296077 & XRT/PC &2009-01-29 15:59:44     &2009-01-29 22:36:24     &5911	\\
00030296078 & XRT/PC &2009-01-30 05:05:45     &2009-01-30 07:01:56     &2477	\\
00030296079 & XRT/PC &2009-01-31 00:23:45     &2009-01-31 02:17:58     &1363    \\
00030296080 & XRT/PC &2009-02-01 19:50:46     &2009-02-01 21:30:57     &2022    \\
00030296081 & XRT/PC &2009-02-02 08:46:12     &2009-02-02 11:40:57     &1886    \\
00030296082 & XRT/PC &2009-02-03 18:25:48     &2009-02-03 21:49:58     &1215    \\
00030296083 & XRT/PC &2009-02-04 00:51:50     &2009-02-04 04:09:56     &1403    \\
00030296084 & XRT/PC &2009-02-05 02:24:25     &2009-02-05 02:50:30     &1507    \\
00030296085 & XRT/PC &2009-02-06 01:13:12     &2009-02-06 04:28:57     &486     \\
00030296087 & XRT/PC &2009-02-08 18:54:08     &2009-02-08 22:15:57     &1605    \\
00030296088 & XRT/PC &2009-02-10 03:10:18     &2009-02-10 08:06:57     &1542    \\
00030296089 & XRT/PC &2009-02-11 15:47:39     &2009-02-11 22:38:57     &2242    \\
00030296090 & XRT/PC &2009-02-12 22:30:18     &2009-02-13 00:20:56     &1683    \\
00030296091 & XRT/PC &2009-02-13 05:04:18     &2009-02-13 06:45:54     &1301    \\
00030296092 & XRT/PC &2009-02-14 06:34:01     &2009-02-14 08:18:56     &1270    \\
00030296093 & XRT/PC &2009-02-15 03:19:41     &2009-02-15 03:39:57     &1202    \\
00030296094 & XRT/PC &2009-02-16 08:26:29     &2009-02-16 10:14:42     &1126    \\
00030296095 & XRT/PC &2009-02-17 00:22:55     &2009-02-17 00:38:50     &952     \\  
 \end{tabular}
  \end{center}
  \end{table*}

\setcounter{table}{5}
 \begin{table*}
 \begin{center}
 \caption{Observation log for IGR~J17391$-$3021.\label{sfxts:tab:alldata17391}}
 \begin{tabular}{llllll}
 \hline
 \noalign{\smallskip}
 Sequence & Instrument/Mode & Start time (UT) &  End time (UT) &  Net Exposure \\
            &     & (yyyy-mm-dd hh:mm:ss) & (yyyy-mm-dd hh:mm:ss) &  (s) \\ 
  \noalign{\smallskip}
 \hline
 \noalign{\smallskip}
 00030987001		 &XRT/PC &2007-10-27 09:46:16	 &2007-10-27 10:05:57	 &	 1181	 \\
 00030987002		 &XRT/PC &2007-10-30 10:12:32	 &2007-10-30 10:25:57	 &	 805	 \\
 00030987003		 &XRT/PC &2007-11-01 08:47:14	 &2007-11-01 09:06:58	 &	 1181	 \\
 00030987004		 &XRT/PC &2008-02-04 21:24:07	 &2008-02-04 22:57:58	 &	 940	 \\
 00030987005		 &XRT/PC &2008-02-06 00:48:04	 &2008-02-06 23:29:58	 &	 5606	 \\
 00030987006	 &XRT/PC &	 2008-02-08 01:10:03	 &	 2008-02-08 02:48:58	 &	 161	 \\
 00030987007	 &XRT/PC &	 2008-02-11 17:26:01	 &	 2008-02-11 19:16:56	 &	 1295	 \\
 00030987008	 &XRT/PC &	 2008-02-13 09:48:13	 &	 2008-02-13 11:30:54	 &	 180	 \\
 00030987009	 &XRT/PC &	 2008-02-15 06:32:34	 &	 2008-02-15 06:40:58	 &	 504	 \\
 00030987010	 &XRT/PC &	 2008-02-18 03:45:06	 &	 2008-02-18 05:26:56	 &	 498	 \\
 00030987011	  &XRT/PC &	  2008-02-21 00:40:01	  &	  2008-02-21 00:57:57	  &	  1076    \\
 00030987012	  &XRT/PC &	  2008-02-22 00:52:08	  &	  2008-02-22 01:06:55	  &	  888	  \\
 00030987013	  &XRT/PC &	  2008-02-25 17:19:52	  &	  2008-02-25 19:03:57	  &	  597	  \\
 00030987014	  &XRT/PC &	  2008-02-27 01:26:01	  &	  2008-02-27 03:05:57	  &	  937	  \\
 00030987015	  &XRT/PC &	  2008-02-28 01:23:19	  &	  2008-02-28 01:40:57	  &	  1056    \\
00030987016    &XRT/PC    &	   2008-03-03 21:06:00     &	   2008-03-03 21:22:58     &	   1016    \\
00030987017    &XRT/PC    &	   2008-03-05 21:21:59     &	   2008-03-05 22:55:57     &	   1244    \\
00030987018    &XRT/PC    &	   2008-03-08 04:03:06     &	   2008-03-08 12:12:57     &	   2821    \\
00030987019    &XRT/PC    &	   2008-03-10 18:43:52     &	   2008-03-10 20:25:56     &	   647     \\
00030987020    &XRT/PC    &	   2008-03-11 12:16:41     &	   2008-03-11 18:54:58     &	   2908    \\
00030987021    &XRT/PC    &	   2008-03-14 09:21:55     &	   2008-03-14 09:40:50     &	   1108    \\
00030987022    &XRT/PC    &	   2008-03-16 17:34:59     &	   2008-03-16 17:51:57     &	   1018    \\
00030987023    &XRT/PC    &	   2008-03-19 00:28:15     &	   2008-03-19 03:45:50     &	   672     \\
00030987024    &XRT/PC    &	   2008-03-21 05:18:20     &	   2008-03-22 00:41:56     &	   1043    \\
00030987025    &XRT/PC    &	   2008-03-23 23:14:57     &	   2008-03-24 20:17:57     &	   1257    \\
00030987026    &XRT/PC    &	   2008-03-30 23:41:20     &	   2008-03-30 23:57:58     &	   996     \\
00030987027    &XRT/PC    &	   2008-04-02 03:08:26     &	   2008-04-02 03:24:56     &	   990     \\
00030987028    &XRT/PC    &	   2008-04-07 16:27:21     &	   2008-04-07 16:43:56     &	   995     \\
00308797000    &BAT/evt   &	   2008-04-08 21:24:16     &	   2008-04-08 23:09:11     &	  1735       \\
00308797000    &XRT/WT    &	   2008-04-08 21:34:47     &	   2008-04-08 23:09:19     &	   128     \\
00308797000    &XRT/PC    &	   2008-04-08 21:36:54     &	   2008-04-08 23:19:19     &	   938     \\
00030987029    &XRT/PC    &	   2008-04-19 06:26:23     &	   2008-04-19 08:09:56     &	   908     \\
00030987030    &XRT/PC    &	   2008-04-21 19:28:21     &	   2008-04-21 21:15:57     &	   1146    \\
00030987031    &XRT/PC    &	   2008-04-23 02:03:03     &	   2008-04-23 02:14:58     &	   689   \\    
00030987032	&XRT/PC   &	  2008-04-28 10:26:57	  &	  2008-04-28 10:42:57	  &	  960	  \\
00030987033	&XRT/PC   &	  2008-04-30 13:58:59	  &	  2008-04-30 14:14:57	  &	  958	  \\
00030987034	&XRT/PC   &	  2008-05-05 16:08:11	  &	  2008-05-05 16:22:56	  &	  885	  \\
00030987035	&XRT/PC   &	  2008-05-07 16:19:15	  &	  2008-05-07 16:34:58	  &	  941	  \\
00030987036	&XRT/PC   &	  2008-05-09 16:27:21	  &	  2008-05-09 16:43:56	  &	  995	  \\
00030987037	&XRT/PC   &	  2008-05-12 08:21:50	  &	  2008-05-12 08:40:56	  &	  1146    \\
00030987038	&XRT/PC   &	  2008-05-14 00:32:06	  &	  2008-05-14 00:51:57	  &	  1191    \\
00030987040	&XRT/PC   &	  2008-05-24 20:51:34	  &	  2008-05-24 21:12:58	  &	  1282    \\
00030987041	&XRT/PC   &	  2008-05-26 22:49:18	  &	  2008-05-26 23:08:56	  &	  1178    \\
00030987042	&XRT/PC   &	  2008-05-28 11:27:22	  &	  2008-05-28 11:45:58	  &	  1116    \\
00030987043	&XRT/PC   &	  2008-05-30 10:18:29	  &	  2008-05-30 10:30:56	  &	  746	  \\
00030987044	&XRT/PC   &	  2008-06-02 16:57:25	  &	  2008-06-02 20:37:56	  &	  1269    \\
00030987045	&XRT/PC   &	  2008-06-04 06:00:05	  &	  2008-06-04 06:18:56	  &	  1131    \\
00030987047	&XRT/PC   &	  2008-06-09 17:54:23	  &	  2008-06-09 18:03:57	  &	  572	  \\
00030987048	&XRT/PC   &	  2008-06-11 06:45:09	  &	  2008-06-11 07:00:56	  &	  948	  \\
00030987049	&XRT/PC   &	  2008-06-12 18:12:07	  &	  2008-06-12 23:07:57	  &	  943	  \\
00030987050	&XRT/PC   &	  2008-06-13 10:04:48	  &	  2008-06-13 10:20:56	  &	  968	  \\
00030987051	&XRT/PC   &	  2008-06-16 00:37:58	  &	  2008-06-16 00:57:56	  &	  1198    \\
00030987052	&XRT/WT   &	  2008-06-21 07:38:11	  &	  2008-06-21 14:00:10	  &	  53	  \\
00030987052	&XRT/PC   &	  2008-06-21 07:38:53	  &	  2008-06-21 14:10:56	  &	  2517    \\
00030987053	&XRT/PC   &	  2008-06-23 09:26:58	  &	  2008-06-23 09:35:58	  &	  538	  \\
00030987054	&XRT/PC   &	  2008-06-25 01:38:49	  &	  2008-06-25 08:02:58	  &	  1863    \\
00030987055	&XRT/PC   &	  2008-07-04 15:29:32	  &	  2008-07-04 15:46:55	  &	  1043    \\
00030987056	&XRT/PC   &	  2008-07-07 10:49:25	  &	  2008-07-07 11:06:48	  &	  1043    \\
00030987057	&XRT/PC   &	  2008-07-09 12:51:21	  &	  2008-07-09 13:02:55	  &	  695	  \\
00030987058	&XRT/PC   &	  2008-07-11 12:41:02	  &	  2008-07-11 12:56:57	  &	  954	  \\
00030987059	&XRT/PC   &	  2008-07-19 05:23:41	  &	  2008-07-19 07:07:57	  &	  992	  \\
00030987060	&XRT/PC   &	  2008-07-21 20:28:02	  &	  2008-07-21 20:34:56	  &	  414	  \\
  \end{tabular}
  \end{center}
  \end{table*}

\setcounter{table}{5}
 \begin{table*}
 \begin{center}
 \caption{Observation log for IGR~J17391$-$3021. Continued.}
 \begin{tabular}{llllll}
 \hline
 \noalign{\smallskip}
 Sequence & Instrument/Mode & Start time (UT) &  End time (UT) &  Net Exposure \\
            &     & (yyyy-mm-dd hh:mm:ss) & (yyyy-mm-dd hh:mm:ss) &  (s) \\ 
  \noalign{\smallskip}
 \hline
 \noalign{\smallskip}
00030987061	&XRT/PC   &	  2008-07-23 17:18:27	  &	  2008-07-23 17:28:09	  &	  579	  \\
00030987062	&XRT/PC   &	  2008-07-25 04:29:36	  &	  2008-07-25 06:14:56	  &	  948	  \\
00030987063	&XRT/PC   &	  2008-07-28 07:58:32	  &	  2008-07-28 09:42:56	  &	  953	  \\
00030987064	&XRT/PC   &	  2008-07-30 19:27:28	  &	  2008-07-30 21:12:55	  &	  1076    \\
00030987067	&XRT/PC   &	  2008-08-06 23:17:02	  &	  2008-08-06 23:38:56	  &	  1314    \\
00030987068	&XRT/PC   &	  2008-08-08 12:09:29	  &	  2008-08-08 12:32:56	  &	  1407    \\
00030987069	&XRT/PC   &	  2008-08-10 02:45:41	  &	  2008-08-10 03:03:57	  &	  1094    \\
00319963000	&XRT/WT   &	  2008-08-13 23:55:55	  &	  2008-08-14 00:29:09	  &	  1688    \\
00319963000	&XRT/PC   &	  2008-08-14 00:03:19	  &	  2008-08-14 00:04:34	  &	  75	  \\
00030987070	&XRT/WT   &	  2008-08-14 01:23:10	  &	  2008-08-14 13:03:50	  &	  1207    \\
00030987070	&XRT/PC   &	  2008-08-14 04:36:11	  &	  2008-08-14 13:17:16	  &	  10714   \\
00030987071	&XRT/PC   &	  2008-08-15 00:00:34	  &	  2008-08-15 00:16:57	  &	  983	  \\
00030987072	&XRT/PC   &	  2008-08-16 03:21:09	  &	  2008-08-16 05:08:57	  &	  1379    \\
00030987073	&BAT/PC   &	  2008-08-17 01:44:09	  &	  2008-08-17 02:01:57	  &	  1068    \\
00030987074	&XRT/PC   &	  2008-08-18 13:04:08	  &	  2008-08-18 13:21:56	  &	  1068    \\
00030987075	&XRT/PC   &	  2008-08-19 13:09:42	  &	  2008-08-19 13:27:58	  &	  1095    \\
00030987076	&XRT/PC   &	  2008-08-20 05:14:06	  &	  2008-08-20 05:31:56	  &	  1071    \\
00030987077	&XRT/PC   &	  2008-08-21 00:30:30	  &	  2008-08-21 15:08:55	  &	  1173    \\
00030987078	&XRT/PC   &	  2008-08-22 00:36:19	  &	  2008-08-22 02:21:55	  &	  1093    \\
00030987079	&XRT/PC   &	  2008-08-23 03:55:11	  &	  2008-08-23 05:41:56	  &	  1229    \\
00030987080	&XRT/PC   &	  2008-08-24 21:44:19	  &	  2008-08-24 23:31:57	  &	  1332    \\
00030987081	&XRT/PC   &	  2008-08-25 21:50:21	  &	  2008-08-25 23:35:56	  &	  1274    \\
00030987082	&XRT/PC   &	  2008-08-27 13:55:09	  &	  2008-08-27 14:15:58	  &	  1249    \\
00030987083	&XRT/PC   &	  2008-08-29 09:16:27	  &	  2008-08-29 11:10:58	  &	  2221    \\
00030987084	&XRT/PC   &	  2008-08-31 14:18:01	  &	  2008-08-31 14:36:57	  &	  1136    \\
00030987085	&BAT/PC   &	  2008-09-04 06:45:52	  &	  2008-09-04 07:06:56	  &	  1263    \\
00030987086	&XRT/PC   &	  2008-09-13 10:44:42	  &	  2008-09-13 23:48:56	  &	  1047    \\
00030987087	&XRT/PC   &	  2008-09-16 22:18:49	  &	  2008-09-16 22:23:35	  &	  286	  \\
00030987088	&XRT/PC   &       2008-09-25 05:31:25	&	2008-09-2505:47:56     &       990 \\
00030987089     &XRT/PC    &       2008-09-29 09:05:25     &       2008-09-29 09:21:56     &       990     \\
00030987091     &XRT/PC    &       2008-10-07 19:33:16     &       2008-10-07 19:50:57     &       1061    \\
00030987092     &XRT/PC    &       2008-10-11 23:08:23     &       2008-10-11 23:21:55     &       812     \\
00030987093     &XRT/PC    &       2008-10-15 18:42:46     &       2008-10-15 20:22:56     &       815     \\
00030987094     &XRT/PC    &       2008-10-19 15:50:55     &       2008-10-19 19:11:56     &       1081    \\
00030987095     &XRT/PC    &       2008-10-23 00:09:33     &       2008-10-23 00:22:58     &       803     \\
00030987096     &XRT/PC    &       2008-10-27 19:50:03     &       2008-10-27 21:33:56     &       875     \\
00030987097     &XRT/PC    &       2008-10-31 00:57:26     &       2008-10-31 01:13:56     &       990     \\
  \end{tabular}
  \end{center}
  \end{table*}

\setcounter{table}{6}
 \begin{table*}
 \begin{center}
 \caption{Observation log for IGR~J17544$-$2619.\label{sfxts:tab:alldata17544}} 
 \begin{tabular}{llllll}
 \hline
 \noalign{\smallskip}
 Sequence & Instrument/Mode & Start time (UT) &  End time (UT) &  Net Exposure \\
            &     & (yyyy-mm-dd hh:mm:ss) & (yyyy-mm-dd hh:mm:ss) &  (s) \\ 
  \noalign{\smallskip}
 \hline
 \noalign{\smallskip}
 00035056002			 &XRT/PC &2007-10-28 00:20:09	 &2007-10-29 07:07:56	 &	 2783	 \\
 00035056003			 &XRT/PC &2007-10-31 10:19:05	 &2007-10-31 13:43:35	 &	 248	 \\
 00035056004			 &XRT/PC &2007-11-04 00:58:32	 &2007-11-04 01:16:58	 &	 1104	 \\
 00035056005			  &XRT/PC &2008-02-07 20:17:58    &2008-02-07 23:43:57    &	  1331    \\
 00035056006	  &XRT/PC &2008-02-10 19:14:07    &	  2008-02-10 20:52:56	  &	  278	  \\
 00035056007	  &XRT/PC &2008-02-14 08:17:17    &	  2008-02-14 10:00:56	  &	  915	  \\
 00035056008	  &XRT/PC &2008-02-17 00:16:44    &	  2008-02-17 02:04:56	  &	  1346    \\
00035056009	 &XRT/PC &	 2008-02-21 10:24:08	 &	 2008-02-21 10:40:36	 &	 960	 \\
00035056010	 &XRT/PC &	 2008-02-24 23:41:09	 &	 2008-02-25 01:24:57	 &	 998	 \\
00035056011	 &XRT/PC &	 2008-02-28 02:41:54	 &	 2008-02-28 02:58:57	 &	 1004	 \\
 00035056012	 &XRT/PC   &2008-03-03 22:43:07       &2008-03-03 23:00:57     &       1071    \\
 00035056013	 &XRT/PC   &2008-03-06 03:52:22       &2008-03-06 07:13:58     &       1180    \\
 00035056014	 &XRT/PC   &2008-03-10 13:54:43       &2008-03-10 17:13:58     &       627     \\
 00035056015	 &XRT/PC   &2008-03-13 22:06:37       &2008-03-13 23:42:56     &       1296    \\
 00035056016	 &XRT/PC   &2008-03-16 04:43:43       &2008-03-16 06:33:56     &       1701    \\
 00035056017	 &XRT/PC   &2008-03-20 12:57:18       &2008-03-20 13:07:58     &       639     \\
 00035056018	 &XRT/PC   &2008-03-23 18:20:14       &2008-03-23 18:36:56     &       1002    \\
 00035056019	 &XRT/PC   &2008-03-27 00:45:04       &2008-03-27 01:02:57     &       1073    \\
 00035056020	 &XRT/PC   &2008-03-31 03:07:59       &2008-03-31 04:52:57     &       898     \\
 00308224000	 & BAT/evt &2008-03-31 20:46:48       &2008-03-31 21:06:50     &       1202   \\
 00308224000	 &XRT/WT   &2008-03-31 20:53:35       &2008-03-31 20:58:57     &       306     \\
 00035056021	 &XRT/PC   &2008-03-31 21:52:49       &2008-03-31 22:17:41     &       1492    \\
 00035056022	 &XRT/PC   &2008-04-02 22:29:22       &2008-04-02 22:42:57     &       815     \\
 00035056024	 &XRT/PC   &2008-04-03 05:05:00       &2008-04-03 06:47:56     &       1098    \\
 00035056025	 &XRT/PC   &2008-04-06 05:12:15       &2008-04-06 05:28:58     &       1000    \\
 00035056026	 &XRT/PC   &2008-04-10 21:46:21     & 2008-04-10 23:30:56     &       1027    \\
 00035056027	 &XRT/PC   &2008-04-13 01:12:20     & 2008-04-13 02:54:28     &       765    \\
 00035056028	 &XRT/PC   &2008-04-17 20:45:21     & 2008-04-17 22:34:57     &       1030    \\
 00035056029	 &XRT/PC   &2008-04-20 16:21:01     & 2008-04-20 19:36:56     &       888     \\		
00035056030	&XRT/PC   &	  2008-04-28 13:44:33	  &	  2008-04-28 14:00:56	  &	  983	  \\
00035056031	&XRT/PC   &	  2008-05-01 19:04:49	  &	  2008-05-01 20:48:56	  &	  925	  \\
00035056032	&XRT/PC   &	  2008-05-04 19:20:41	  &	  2008-05-04 19:30:58	  &	  615	  \\
00035056033	&XRT/PC   &	  2008-05-11 16:18:15	  &	  2008-05-11 16:35:58	  &	  1061    \\
00035056035	&XRT/PC   &	  2008-05-29 11:56:34	  &	  2008-05-29 13:40:56	  &	  1122    \\
00035056036	&XRT/PC   &	  2008-06-01 01:05:55	  &	  2008-06-01 02:46:56	  &	  1091    \\
00035056037	&XRT/PC   &	  2008-06-08 21:01:29	  &	  2008-06-08 21:12:56	  &	  687	  \\
00035056038	&XRT/PC   &	  2008-06-12 19:49:25	  &	  2008-06-12 21:35:57	  &	  1269    \\
00035056040	&XRT/PC   &	  2008-06-22 16:04:17	  &	  2008-06-22 17:49:56	  &	  1098    \\
00035056041	&XRT/PC   &	  2008-06-26 10:03:44	  &	  2008-06-26 11:48:57	  &	  1274    \\
00035056042	&XRT/PC   &	  2008-06-29 00:48:45	  &	  2008-06-29 02:27:57	  &	  607	  \\
00035056043	&XRT/PC   &	  2008-07-03 13:54:30	  &	  2008-07-03 15:37:57	  &	  1071    \\
00035056044	&XRT/PC   &	  2008-07-06 18:50:24	  &	  2008-07-06 19:10:57	  &	  1232    \\
00035056045	&XRT/PC   &	  2008-07-10 12:48:28	  &	  2008-07-10 12:52:21	  &	  232	  \\
00035056046	&XRT/PC   &	  2008-07-13 13:03:40	  &	  2008-07-13 14:47:56	  &	  945	  \\
00035056047	&XRT/PC   &	  2008-07-18 00:57:56	  &	  2008-07-18 02:10:56	  &	  900	  \\
00035056048	&XRT/PC   &	  2008-07-20 10:33:42	  &	  2008-07-20 10:52:56	  &	  1153    \\
00035056049	&XRT/PC   &	  2008-07-24 12:34:50	  &	  2008-07-24 12:37:33	  &	  163	  \\
00035056051	&XRT/PC   &	  2008-07-31 01:46:30	  &	  2008-07-31 02:02:56	  &	  985	  \\
00035056052	&XRT/PC   &	  2008-08-03 10:14:34	  &	  2008-08-03 10:34:57	  &	  1222    \\
00035056053	&XRT/PC   &	  2008-08-07 15:10:57	  &	  2008-08-07 15:14:56	  &	  238	  \\
00035056054	&XRT/PC   &	  2008-08-10 04:22:25	  &	  2008-08-10 05:55:58	  &	  1320    \\
00035056055	&XRT/PC   &	  2008-08-14 20:46:21	  &	  2008-08-14 21:02:57	  &	  994	  \\
00035056056	&XRT/PC   &	  2008-08-17 03:20:43	  &	  2008-08-17 03:38:56	  &	  1093    \\
00035056057	&XRT/PC   &	  2008-08-21 16:34:57	  &	  2008-08-21 18:15:55	  &	  1015    \\
00035056058	&XRT/PC   &	  2008-08-24 18:37:32	  &	  2008-08-24 18:53:55	  &	  983	  \\
00035056059	&XRT/PC   &	  2008-08-29 23:52:42	  &	  2008-08-30 05:05:58	  &	  993	  \\
00035056060	&XRT/PC   &	  2008-08-31 15:57:58	  &	  2008-08-31 22:25:57	  &	  1191    \\
 \end{tabular}
  \end{center}
  \end{table*}

\setcounter{table}{6}
 \begin{table*}
 \begin{center}
 \caption{Observation log for IGR~J17544$-$2619. Continued.} 
 \begin{tabular}{llllll}
 \hline
 \noalign{\smallskip}
 Sequence & Instrument/Mode & Start time (UT) &  End time (UT) &  Net Exposure \\
            &     & (yyyy-mm-dd hh:mm:ss) & (yyyy-mm-dd hh:mm:ss) &  (s) \\ 
  \noalign{\smallskip}
 \hline
 \noalign{\smallskip}
00035056061	&XRT/PC   &	  2008-09-04 00:12:45	  &	  2008-09-04 00:26:55	  &	  632	  \\
00035056061	&XRT/WT   &	  2008-09-04 00:12:40	  &	  2008-09-04 00:25:03	  &	  217	  \\
00035056062	&XRT/PC   &	  2008-09-05 05:22:27	  &	  2008-09-05 11:44:56	  &	  2521    \\
00035056065	&XRT/PC   &	  2008-09-11 09:05:46	  &	  2008-09-11 10:44:55	  &	  923	  \\
00035056066	&XRT/PC   &	  2008-09-12 12:28:46	  &	  2008-09-12 12:44:56	  &	  970	  \\
00035056067	&XRT/PC   &	  2008-09-13 12:23:36	  &	  2008-09-13 14:13:56	  &	  900	  \\
00035056068	&XRT/PC   &	  2008-09-14 12:31:39	  &	  2008-09-14 14:18:55	  &	  905	  \\
00035056069	&XRT/PC   &	  2008-09-15 10:54:20	  &	  2008-09-15 12:45:55	  &	  958	  \\
00035056070	&XRT/PC   &	  2008-09-16 22:12:44	  &	  2008-09-16 23:59:57	  &	  930	  \\
00035056071	&XRT/PC   &	  2008-09-19 00:03:19	  &	  2008-09-19 00:04:34	  &	  75	  \\
00035056072	&XRT/PC   &	  2008-09-22 02:02:35	  &	  2008-09-22 02:24:56	  &	  1341    \\
00035056073	&XRT/PC   &	  2008-09-26 13:39:41	  &	  2008-09-26 13:48:27	  &	  527 \\
00035056074     &XRT/PC   &	  2008-09-30 07:33:35	  &	  2008-09-30 07:50:55	  &	  1041    \\
00035056075     &XRT/PC   &	  2008-10-04 01:41:09	  &	  2008-10-04 01:55:56	  &	  888	  \\
00035056076     &XRT/PC   &	  2008-10-08 11:37:00	  &	  2008-10-08 14:55:56	  &	  1248    \\
00035056077     &XRT/PC   &	  2008-10-12 20:01:34	  &	  2008-10-12 21:43:57	  &	  644	  \\
00035056078     &XRT/PC   &	  2008-10-16 06:21:58	  &	  2008-10-16 06:35:58	  &	  839	  \\
00035056079     &XRT/PC   &	  2008-10-20 09:30:55	  &	  2008-10-20 11:20:58	  &	  1683    \\
00035056080     &XRT/PC   &	  2008-10-25 05:11:59	  &	  2008-10-25 06:58:56	  &	  1249    \\
00035056081     &XRT/PC   &	  2008-10-28 07:08:43	  &	  2008-10-28 10:23:57	  &	  506	  \\
00035056082     &XRT/PC   &	  2008-10-31 01:14:50	  &	  2008-10-31 01:30:58	  &	  967	  \\
 \end{tabular}
  \end{center}
  \end{table*}

\setcounter{table}{7}

 \begin{table*}
 \begin{center}
 \caption{Observation log for IGR~J18410$-$0535.\label{sfxts:tab:alldata18410}} 
 \begin{tabular}{llllll}
 \hline
 \noalign{\smallskip}
 Sequence & Instrument/Mode & Start time (UT) &  End time (UT) &  Net Exposure \\
            &     & (yyyy-mm-dd hh:mm:ss) & (yyyy-mm-dd hh:mm:ss) &  (s) \\ 
  \noalign{\smallskip}
 \hline
 \noalign{\smallskip}
00030988001	&XRT/PC	&2007-10-26 00:08:53	&2007-10-26 06:45:56	&	1384	\\
00030988002	&XRT/PC &2007-10-28 22:53:11	&2007-10-29 23:12:56	&	3199	\\
00030988003	&XRT/PC &2007-11-03 10:37:52	&2007-11-03 12:24:56	&	1122	\\
00030988004	&XRT/PC &2007-11-05 09:08:39	&2007-11-05 09:28:58	&	1218	\\
00030988005	&XRT/PC &2007-11-09 16:11:30	&2007-11-09 16:32:56	&	1286  \\
00030988006	&XRT/PC &2007-11-12 14:51:25	&2007-11-12 14:51:55	&	30	\\
00030988007	&XRT/PC &2007-11-16 08:40:04	&2007-11-16 10:36:58	&	2332	\\
00030988008	&XRT/PC &2007-11-18 23:22:32	&2007-11-18 23:41:58	&	1165	\\
00030988009	&XRT/PC &	2008-02-14 14:44:51	&	2008-02-14 16:28:58	&	947	\\
00030988009	&XRT/WT &	2008-02-14 14:42:45	&	2008-02-14 16:20:15	&	127	\\
00030988010	&XRT/PC &	2008-02-18 00:26:32	&	2008-02-18 02:13:57	&	626	\\
00030988011	&XRT/PC &	2008-02-21 08:46:13	&	2008-02-21 09:02:58	&	1005	\\
00030988012	&XRT/PC &	2008-02-25 20:32:34	&	2008-02-25 20:47:57	&	834	\\
00030988013	&XRT/PC &	2008-02-28 03:00:08	&	2008-02-28 03:16:56	&	1008\\
00030988014	&XRT/PC    &	   2008-03-03 17:54:04     &	   2008-03-03 17:56:45     &	   160     \\
00030988015	&XRT/PC    &	   2008-03-06 08:41:40     &	   2008-03-06 12:00:57     &	   1374    \\
00030988016	&XRT/PC    &	   2008-03-10 10:49:43     &	   2008-03-10 23:20:57     &	   2031    \\
00030988017	&XRT/PC    &	   2008-03-13 05:59:01     &	   2008-03-13 06:20:58     &	   1316    \\
00030988018	&XRT/PC    &	   2008-03-17 00:12:23     &	   2008-03-17 01:54:57     &	   1143    \\
00030988019	&XRT/PC    &	   2008-03-20 17:56:07     &	   2008-03-20 18:16:58     &	   1249    \\
00030988020	&XRT/PC    &	   2008-03-24 02:22:04     &	   2008-03-24 04:07:56     &	   1173    \\
00030988021	&XRT/PC    &	   2008-03-27 01:04:11     &	   2008-03-27 01:19:56     &	   944     \\
00030988022	&XRT/PC    &	   2008-03-31 01:25:11     &	   2008-03-31 01:41:57     &	   1005    \\
00030988023	&XRT/PC    &	   2008-04-03 08:17:07     &	   2008-04-03 10:02:49     &	   973     \\
00030988024	&XRT/PC    &	   2008-04-07 05:18:10     &	   2008-04-07 05:34:56     &	   1005    \\
00030988026	&XRT/PC    &	   2008-04-17 12:52:27     &	   2008-04-17 14:35:56     &	   1078    \\
00030988027	&XRT/PC    &	   2008-04-21 08:14:08     &	   2008-04-21 08:31:56     &	   1068    \\
00030988029	&XRT/PC    &	   2008-04-28 08:51:35     &	   2008-04-28 09:07:56     &	   980     \\
00030988030	&XRT/PC    &	   2008-05-01 15:53:55     &	   2008-05-01 17:37:56     &	   845     \\
00030988031	&XRT/PC    &	   2008-05-05 12:54:44     &	   2008-05-05 14:44:58     &	   925     \\
00030988032	&XRT/PC    &	   2008-05-08 03:10:56     &	   2008-05-08 05:23:57     &	   888     \\
00030988033	&XRT/PC    &	   2008-05-12 01:57:30     &	   2008-05-12 02:14:58     &	   1048    \\
00030988034	&XRT/PC    &	   2008-05-15 12:14:47     &	   2008-05-15 13:54:56     &	   1081    \\
00030988035	&XRT/PC    &	   2008-05-22 03:22:44     &	   2008-05-22 03:29:58     &	   434     \\
00030988036	&XRT/PC    &	   2008-05-26 19:23:27     &	   2008-05-26 19:42:58     &	   1171    \\
00030988037	&XRT/PC    &	   2008-05-29 00:43:05     &	   2008-05-29 19:59:58     &	   1402    \\
00030988038	&XRT/PC    &	   2008-06-02 12:01:25     &	   2008-06-02 12:17:58     &	   991     \\
00030988039	&XRT/PC    &	   2008-06-01 07:37:34     &	   2008-06-02 20:21:56     &	   1263    \\
00030988041	&XRT/PC    &	   2008-06-11 13:03:36     &	   2008-06-11 13:18:56     &	   920     \\
00030988042	&XRT/PC    &	   2008-06-18 07:26:24     &	   2008-06-18 07:46:57     &	   1234    \\
00030988043	&XRT/PC    &	   2008-06-21 07:28:07     &	   2008-06-21 13:57:58     &	   2188    \\
00030988044	&XRT/PC    &	   2008-06-23 04:28:01     &	   2008-06-23 07:46:55     &	   1301    \\
00030988045	&XRT/PC    &	   2008-06-25 01:57:19     &	   2008-06-25 08:18:55     &	   1123    \\
00030988046	&XRT/PC    &	   2008-06-28 13:13:57     &	   2008-06-28 13:30:57     &	   1020    \\
00030988047	&XRT/PC    &	   2008-06-30 21:50:01     &	   2008-06-30 21:50:56     &	   54	   \\
00030988048	&XRT/PC    &	   2008-07-04 13:32:01     &	   2008-07-04 13:49:57     &	   1075    \\
00030988049	&XRT/PC    &	   2008-07-07 12:13:17     &	   2008-07-07 12:30:55     &	   1058    \\
00030988050	&XRT/PC    &	   2008-07-09 19:03:33     &	   2008-07-09 19:06:56     &	   203     \\
00030988051	&XRT/PC    &	   2008-07-11 00:16:45     &	   2008-07-11 01:59:56     &	   604     \\
00030988052	&XRT/PC    &	   2008-07-14 13:09:38     &	   2008-07-14 16:27:57     &	   963     \\
00030988053	&XRT/PC    &	   2008-07-16 00:42:49     &	   2008-07-16 02:31:57     &	   1214    \\
00030988054	&XRT/PC    &	   2008-07-18 18:17:53     &	   2008-07-18 18:36:56     &	   1143    \\
00030988055	&XRT/PC    &	   2008-07-21 22:04:43     &	   2008-07-21 22:16:58     &	   734     \\
00030988056	&XRT/PC    &	   2008-07-23 19:15:35     &	   2008-07-23 19:15:58     &	   21	   \\
00030988057	&XRT/PC    &	   2008-07-25 00:02:37     &	   2008-07-25 01:42:58     &	   1088    \\
00030988058	&XRT/PC    &	   2008-07-28 06:32:48     &	   2008-07-28 06:35:56     &	   187     \\
00030988059	&XRT/PC    &	   2008-07-30 11:38:47     &	   2008-07-30 11:52:57     &	   850     \\
00030988059	&XRT/WT    &	   2008-07-30 11:38:14     &	   2008-07-30 11:38:45     &	   31	   \\
00030988060	&XRT/PC    &	   2008-08-01 18:05:50     &	   2008-08-01 18:21:58     &	   968     \\
  \end{tabular}
  \end{center}
  \end{table*}

\setcounter{table}{7}
 \begin{table*}
 \begin{center}
 \caption{Observation log for IGR~J18410$-$0535. Continued} 
 \begin{tabular}{llllll}
 \hline
 \noalign{\smallskip}
 Sequence & Instrument/Mode & Start time (UT) &  End time (UT) &  Net Exposure \\
            &     & (yyyy-mm-dd hh:mm:ss) & (yyyy-mm-dd hh:mm:ss) &  (s) \\ 
  \noalign{\smallskip}
 \hline
 \noalign{\smallskip}
00030988061	&XRT/PC    &	   2008-08-03 15:26:25     &	   2008-08-04 20:08:57     &	   1051    \\
00030988062	&XRT/PC    &	   2008-08-06 18:32:32     &	   2008-08-06 18:49:55     &	   1043    \\
00030988064	&XRT/PC    &	   2008-08-11 12:25:28     &	   2008-08-11 12:40:10     &	   883     \\
00030988065	&XRT/PC    &	   2008-08-16 08:05:17     &	   2008-08-16 08:21:58     &	   1000    \\
00030988066	&XRT/PC    &	   2008-08-18 10:10:52     &	   2008-08-18 10:26:57     &	   965     \\
00030988067	&XRT/PC    &	   2008-08-20 06:51:23     &	   2008-08-20 07:07:56     &	   993     \\
00030988068	&XRT/PC    &	   2008-08-22 03:50:07     &	   2008-08-22 05:32:56     &	   930     \\
00030988069	&XRT/PC    &	   2008-08-25 07:32:25     &	   2008-08-25 08:00:57     &	   1712    \\
00030988070	&XRT/PC    &	   2008-08-27 12:19:48     &	   2008-08-27 12:38:56     &	   1148    \\
00030988071	&XRT/PC    &	   2008-08-30 06:24:02     &	   2008-08-30 06:39:57     &	   955     \\
00030988072	&XRT/PC    &	   2008-08-31 23:58:41     &	   2008-09-01 04:57:57     &	   2579    \\
00030988073	&XRT/PC    &	   2008-09-03 00:10:17     &	   2008-09-03 03:47:57     &	   1224    \\
00030988074	&XRT/PC    &	   2008-09-05 13:18:16     &	   2008-09-05 13:30:56     &	   760     \\
00030988075	&XRT/PC    &	   2008-09-08 10:22:05     &	   2008-09-08 12:09:56     &	   1379    \\
00030988076	&XRT/PC    &	   2008-09-12 14:01:24     &	   2008-09-12 14:17:57     &	   993     \\
00030988077	&XRT/PC    &	   2008-09-15 19:18:47     &	   2008-09-15 19:33:57     &	   910     \\
00030988078	&XRT/PC    &	   2008-09-17 08:03:00     &	   2008-09-17 11:28:58     &	   988     \\
00030988079	&XRT/PC    &	   2008-09-24 18:33:17	  &	  2008-09-24 20:19:57	  &	  1399    \\
00030988080	&XRT/PC    &	   2008-09-28 17:25:46	  &	  2008-09-28 19:09:56	  &	  1106    \\
00030988081     &XRT/PC    &	   2008-10-02 17:31:43     &	   2008-10-02 17:48:56     &	   1033    \\
00030988082     &XRT/PC    &	   2008-10-06 00:11:19     &	   2008-10-06 00:28:58     &	   984     \\
00030988083     &XRT/PC    &	   2008-10-10 13:41:49     &	   2008-10-10 13:57:56     &	   968     \\
00030988085     &XRT/PC    &	   2008-10-17 04:58:39     &	   2008-10-17 06:30:56     &	   1632    \\
00030988086     &XRT/PC    &	   2008-10-22 20:57:48     &	   2008-10-22 22:46:57     &	   1582    \\
00030988087     &XRT/PC    &	   2008-10-26 00:30:59     &	   2008-10-26 08:40:12     &	   940     \\
00030988088     &XRT/PC    &	   2008-10-30 21:44:03     &	   2008-10-30 22:02:56     &	   1133    \\
00030988089     &XRT/PC    &	   2008-11-02 01:30:42     &	   2008-11-02 01:49:56     &	   1153    \\
00030988090     &XRT/PC    &	   2008-11-08 01:42:41     &	   2008-11-08 03:29:57     &	   1324    \\
00030988091     &XRT/PC    &	   2008-11-12 05:29:21     &	   2008-11-12 08:52:56     &	   2091    \\
00030988092     &XRT/PC    &	   2008-11-15 08:50:09     &	   2008-11-15 09:11:58     &	   1309    \\
  \end{tabular}
  \end{center}
  \end{table*}

\bsp

\label{lastpage}

\end{document}